\documentclass[sigplan,10pt]{acmart}
\AtBeginDocument{%
  }

\acmYear{2026}\copyrightyear{2026}
\setcopyright{cc}
\setcctype[4.0]{by}
\acmConference[EUROSYS '26]{European Conference on Computer Systems}{April 27--30, 2026}{Edinburgh, Scotland Uk}
\acmBooktitle{European Conference on Computer Systems (EUROSYS '26), April 27--30, 2026, Edinburgh, Scotland Uk}
\acmDOI{10.1145/3767295.3769391}
\acmISBN{979-8-4007-2212-7/26/04}
\acmSubmissionID{15281.77}

\usepackage[T1]{fontenc}   % T1 encoding (required)
\usepackage{amsthm}
\usepackage{pifont}
\newcommand{\cmark}{\ding{51}} % define checkmark command
\newcommand{\xmark}{\ding{55}} % define crossmark command
\usepackage{multirow}
\usepackage{amsfonts}
\usepackage{xspace}
\usepackage[noend]{algorithm2e}
\newcommand{\tc}{$\tau_c$\xspace}
\newcommand{\mtc}{\tau_c} 
\usepackage{subfigure}
\usepackage{listings}
\usepackage{soul}
\begin{document}

\title{2DIO: Configurable and Cache-Accurate Trace
Generation for Storage Benchmarking}

\author{Yirong Wang}
\orcid{0009-0004-6303-5435}
\affiliation{%
  \institution{Northeastern University}
  \city{Boston}
  \state{Massachusetts}
  \country{USA}
}
\email{wang.yiron@northeastern.edu}

\author{Isaac Khor}
\orcid{0009-0005-7917-0292}
\affiliation{%
  \institution{Northeastern University}
  \city{Boston}
  \state{Massachusetts}
  \country{USA}
}
\email{khor.i@northeastern.edu}

\author{Peter Desnoyers}
\orcid{0000-0002-6194-2806}
\affiliation{%
  \institution{Northeastern University}
  \city{Boston}
  \state{Massachusetts}
  \country{USA}
}
\email{P.Desnoyers@northeastern.edu}

\begin{abstract}
  We introduce 2DIO, a microbenchmark creating cache-accurate, stressful I/O traces. While existing tools are limited to generating traces with well-behaved, \emph{concave} hit ratio curves, 2DIO produces ones with tunable complex cache behaviors, particularly performance \emph{cliffs} and \emph{plateaus}.

Our framework encodes a workload as a compact parameter triplet, capturing both short‑term recency and long‑term frequency. This parsimonious parameterization allows researchers to easily translate individual adjustments into predictable cache effects across various eviction policies, and enables the parameter space to be "swept" for exhaustive exploration of desired cache behavior, or to mimic real traces by calibrating parameters to match observed behaviors. 

The tuned parameters are portable, meaning
if the scale of the system under evaluation changes, so too will the footprint and length of the trace, while the relative cache behaviors are preserved.

Evaluations demonstrate 2DIO's ability to generate traces across a continuum of "what-if" cache behaviors and to reproduce real-world ones with high accuracy.
\end{abstract}

  


% \received{16 May 2025}
% \received[revised]{24 September 2025}
% \received[accepted]{26 September 2025}

% \settopmatter{printfolios=false}
\maketitle

%-------------------------------------------------------------------------------
\section{Introduction}
\label{sec:intro}
%------------------------------------------------------------------------------- 

Storage system research requires both measuring the performance of storage systems, and comparing these measured results against other systems.
These systems are often large and complex, used by complex applications.
Performance is affected not only by system hardware and configuration, but by application or workload characteristics, so measurements are inherently tied to the workload used when making them.
Storage systems exhibit complex workload-dependent behaviors, and thus the choice of workload is key to measurements which can be used to compare and predict performance of these systems.

In an ideal world, these  measurements would be: (1) predictive, allowing us to accurately estimate the performance of a system (or the relative performance of different systems) in future executions;
(2) generalizable, providing predictions (especially comparative ones) which hold over as wide a range of applications and configurations as possible;
(3) reproducible, i.e. different researchers should be able to make the same measurement and compare results, and
(4) easy to perform, at an effort and cost which is not excessive relative to the system under test.

\subsection{Benchmark Types}
We begin by reviewing three ways to evaluate storage: (a) real workloads, i.e. the target application itself; (b) trace replay, recorded operations performed under a real workload and reproducing them with more or less fidelity to the original, and (c) synthetic workloads, with algorithmically-generated behaviors intended to mimic either real workloads or some aspect of their behaviors.

\noindent \textbf{Real applications.} In certain cases, e.g. HPC procurement, real application benchmarks are the ``gold standard'', giving the exact answer needed; however they pose a number of deficiencies for storage research.
In particular results are not generalizable; if user A tests a system on workload A, while user B tests a different one on workload B, little is learned about system A vs B.

\noindent \textbf{Trace replay.} Trace replay offers a way to test different systems with identical, real-world workloads.
Individual I/O operations  are recorded on live systems, with trace corpuses like {CloudPhysics}\cite{waldspurger_efficient_2015} and {AliCloud}~\cite{alibaba} offering diverse workloads.
Standard (e.g. {fio}~\cite{fio}) or custom replay tools are used to replay these traces against a system under test, allowing the original workload to be repeated on demand.
But modern trace corpuses are large and cumbersome (70\,GB and 700\,GB for  {CloudPhysics} and {AliCloud}), making exhaustive replay impractical, and new corpuses are rare due to high procedural barriers to releasing potentially sensitive data, leading to long delays before new applications (e.g. LLMs) are reflected in available traces.

\noindent \textbf{Synthetic traces.} 
Synthetic trace generators that mimic real workloads address many issues with trace replay. Without
the need for huge trace libraries, synthetic generators can be readily integrated into many tools. 

Conceptually the task of creating such a generator is straightforward: measure the workload, encode it in a probabilistic model, and generate new traces from that model. With parameters capturing the key workload features, one can systematically explore the entire space rather than selecting arbitrary points from a set of opaque real traces.

But what characteristics to measure and reproduce? The answer seems simple: those that affect the system under test. If performance depends on I/O size, record and reproduce its distribution; if it differs between reads and writes, capture that ratio; if it varies with seek distance, model that as well. 

If the system being evaluated is a cache, one would argue that the hit ratio curve (HRC), i.e., the hit ratio as a function of cache size, is the most critical. Although generators such as fio support frequency-skew models (Zipf, Pareto, Zoned) that influence the HRC, we show that this alone is far from sufficient to reproduce the HRCs observed in real workloads.

\subsection{Cache-accurate Synthetic Traces}
\label{sec:intro:ird}
Workload generators are typically not used to predict the behavior of systems, but rather to compare them. An example is the evaluation sections of papers published in this conference, where data storage papers frequently use fio, Filebench~\cite{Filebench}, and trace replay to compare their system to prior work. These evaluations use (a) reproducible workloads, allowing experiments to be compared, (b) multiple workloads covering a range of possible application behaviors, and (c) at least some realistic workloads, mimicking real application behaviors. If system A out-performs B on most or all the tests in a properly-done evaluation, the community tends to believe this indicates a high probability that A will out-perform B in the field.

For systems where performance is heavily influenced by cache hit rate, the cacheability of this workload is important, as otherwise our benchmarks may measure miss performance when hits would be seen in the field, or vice versa.\footnote{This was an issue in the IRCache web cache ``bake-offs’’ several decades ago~\cite{ircache_2000}, where the use of uniform randomly-distributed workloads disadvantaged some vendors who had used more advanced caching algorithms.}

\noindent\textbf{Frequency distributions.}
The state of the art in reproducing the cacheability of workload appears to be item \emph{frequency} models, using either a parameterized distribution (typically Zipf) or empirically-measured ones from existing workloads \cite{wang2022separating,pletka2018management,stoica2019understanding,kang20202r, wang2020cache, tan2019ibtune, yuan2024csea}.
The origin of this approach is Denning's \emph{Independent Reference Model} (IRM)~\cite{aho1971principles}: each arrival is assumed to be an independently weighted choice from the set of possible items.

% Typical microbenchmarks \cite{iozone, fio, bonnie, iometer, sysbench, Filebench} model workloads primarily through a few coarse knobs---read/write ratio, object size, or broader "workload personalities" such as database, file, or web servers. The cache sees whatever spatio-temporal locality naturally arises from those aggregate knobs, making the workloads effectively cache-agnostic at block level.
% Among these tools, some~\cite{fio,iometer,sysbench} offer IRM block frequency specifications, e.g., Zipf, Pareto, or customized Zoned distributions. 
\begin{figure}
    \centering 
    \subfigure[IRD histogram]{
        \includegraphics[width=0.47\columnwidth]{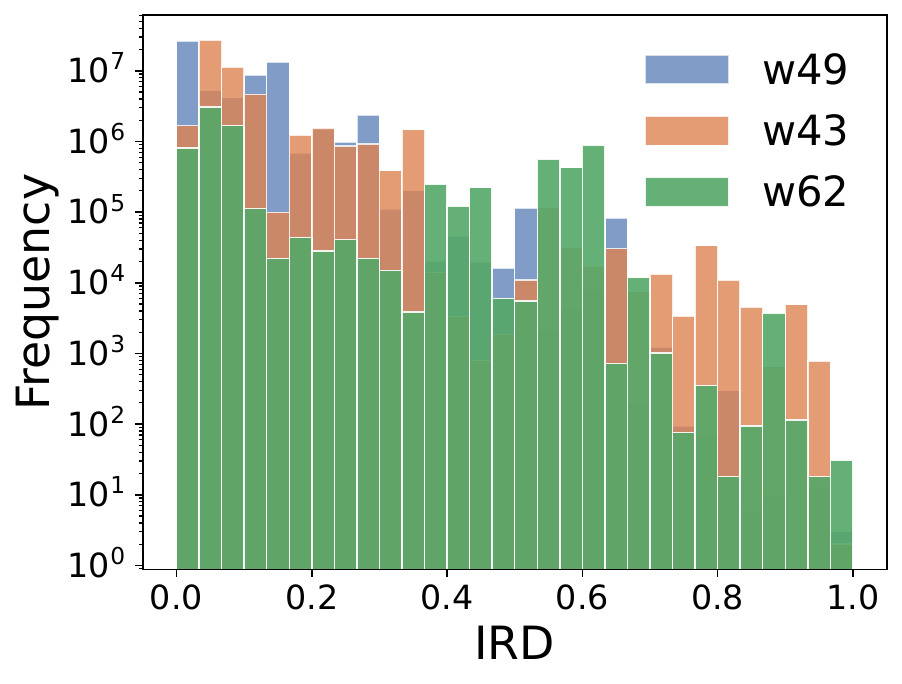}
        \label{fig:w43:IRD}
    } \hfill
    \subfigure[LRU hit ratio]{
        \includegraphics[width=0.46\columnwidth]{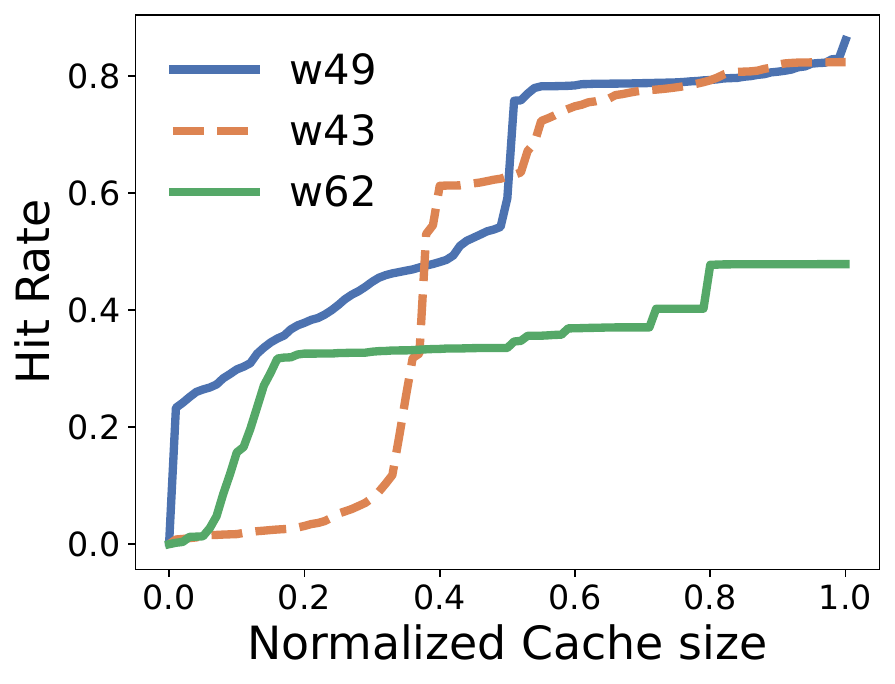}
        \label{fig:lru:3traces}
      }
    \caption{Several CloudPhysics traces showing diverse hit rate behavior. Cache size is normalized to the trace footprint, i.e. the total number of unique blocks accessed in the trace.}
   \label{fig:real_ird_HRC}
   \centering
    \subfigure[IRD histogram]{
        \includegraphics[width=0.47\columnwidth]{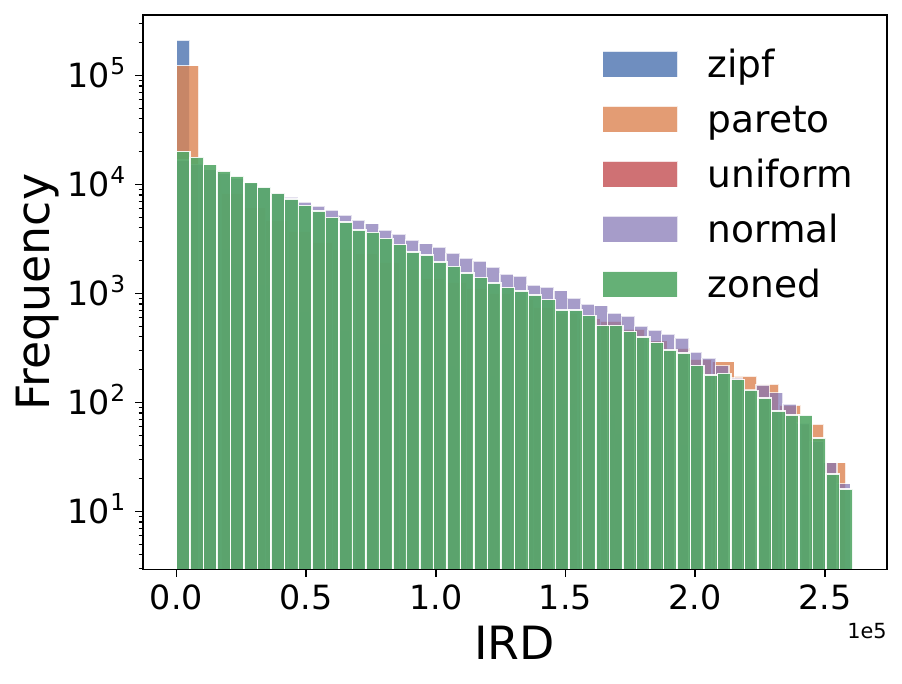}
        \label{fig:irm:IRD}
    } \hfill
    \subfigure[LRU hit ratio]{
        \includegraphics[width=0.47\columnwidth]{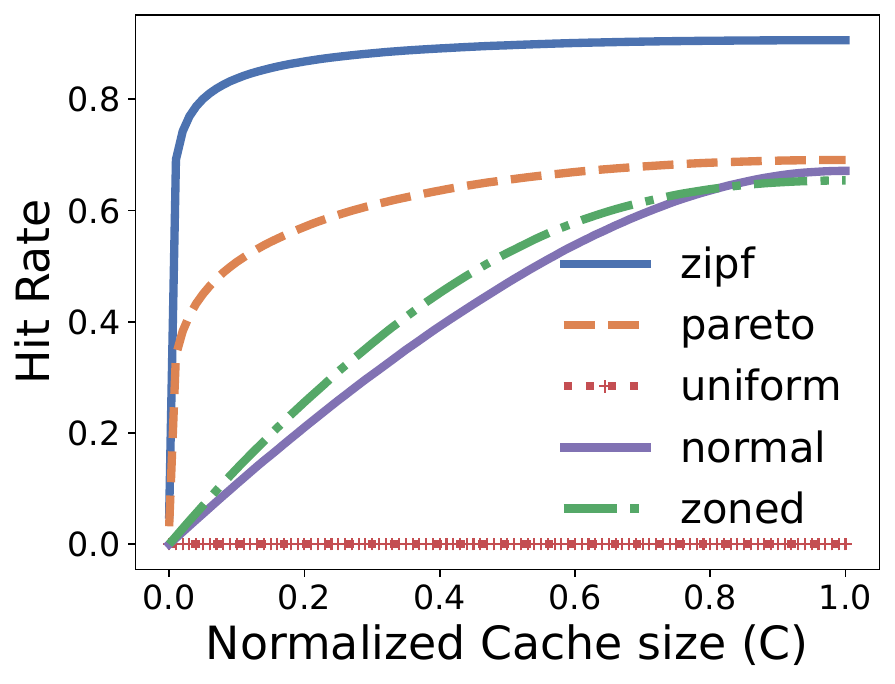}
        \label{fig:irm:LRU}
    } 
    \caption{IRD distributions and hit rate behavior for fio-generated synthetic traces.}
    \label{fig:complex_vs_simple_IRD}
    \vspace{-0.2in}
  \end{figure}  

IRM can be accurate for e.g., content delivery networks (CDNs) where many independent sources are aggregated into a single stream, yet it misses the mark wildly for block storage systems, where typical workloads are comprised of a few highly correlated streams.
For example, Fig.~\ref{fig:real_ird_HRC} demonstrates the behavior of three real-world block traces from the CloudPhysics corpus~\cite{waldspurger_efficient_2015, waldspurger_cache_2017}. Their inter-reference distance (IRD) histograms (Fig.~\ref{fig:w43:IRD}) feature prominent \emph{spikes} and \emph{holes}. When fed into a Least-Recently-Used (LRU) cache, these traces produce complex \emph{non-concave} HRCs (Fig.~\ref{fig:lru:3traces}), including performance \emph{cliffs}, where small cache increases lead to large gains, and \emph{plateaus}, where additional cache provides little benefit.
In contrast, Fig.~\ref{fig:complex_vs_simple_IRD} shows fio-generated workloads, whose IRD distributions are strictly decreasing, yielding \emph{concave} HRCs, reflecting diminishing returns from added cache.

\noindent\textbf{Recency distributions.}
\emph{It is mathematically impossible to manipulate frequency distributions in a way that would produce \emph{non-concave} LRU HRCs~\cite{van1993properties}}.
Whether an access will hit in an LRU cache is solely determined by the IRD since the preceding access to the same item, and, in particular, whether it was evicted during this interval. Since IRM has an equal probability of accessing a particular item at each reference, these IRDs will always be exponentially distributed, while non-concavity in HRC arises precisely because IRDs (e.g. under scan-like workloads) are not memoryless.

The cache modeling community~\cite{eklov2010statstack, niu2012parda, sen2013reuse, xiang2011linear, hu2015lama, counterstacks2014, hao_che_hierarchical_2002, waldspurger_cache_2017, waldspurger_efficient_2015, hasslinger_scope_2023} has long been using \emph{recency} models such as {stack distance} (SD)~\cite{mattson_evaluation_1970} and IRD~\cite{denning_working_1968,denning1972properties} to characterize non-memoryless traffic. Despite their effectiveness in predicting the HRC of a given trace, we are not aware of any prior work using them in the reverse direction, i.e., generating trace(s) matching a target HRC.

To fill this gap, we introduce {Gen-from-IRD}, an algorithm that generates accesses based on a given IRD distribution. Traces generated via {Gen-from-IRD} precisely exhibit a target LRU HRC, and are far more accurate than IRM-generated traces for cache algorithms that use recency.

\noindent\textbf{Frequency + recency distributions.}
We present 2DIO, a synthetic trace generator that faithfully produces (or reproduces) access frequency and recency. 

Building on Gen‑from‑IRD, 2DIO encodes these characteristics into a compact parameter triplet, the \emph{trace profile}, which creates or “counterfeit” workloads exhibiting desired performance \emph{cliffs}/\emph{plateaus}. Researchers can thus customize traces with precise target HRCs or mimic a real workload at various scales for diverse benchmarking tasks.

\begin{figure}[t!]
    \centering
    \includegraphics[width=0.95\columnwidth]{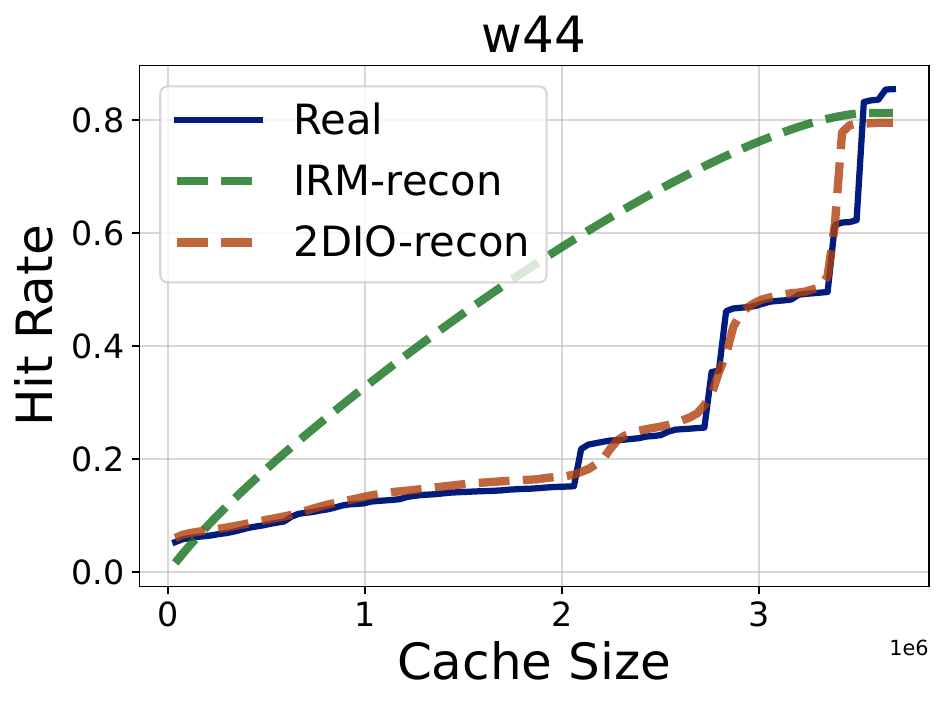}
    \caption{LRU HRCs for {CloudPhysics} {w44}: original trace (blue), 2DIO-generated trace (orange), and trace reconstructed using empirically-measured item frequency distribution (green). Cache size is measured in number of blocks.}
    \label{fig:simple_HRC}
   
    \end{figure}

2DIO does this by merging recency (IRD) models for short-term behavior with frequency (IRM) models for long-term behavior. Its effectiveness is shown in Fig.~\ref{fig:simple_HRC}, which displays the LRU HRC of {CloudPhysics w44}, as well as IRM- and 2DIO-reconstructions based on empirically measured characteristics.
Although the IRM-reconstruction faithfully reproduces the access frequencies for different blocks in the trace, the result is a simple, concave HRC, with none of the performance anomalies seen with the original trace being reproduced.
In contrast, the 2DIO-generated trace yields results very similar to the original, reproducing the performance \emph{cliffs} and \emph{plateaus} with only minor deviations.

This approach applies to broader areas, e.g., CDNs and web caches, where performance is measured by object or request hit rate regardless of object size.

Succinct parameterization is essential for tractability.
For this purpose, 2DIO (a) approximates arbitrary non-memoryless IRD distribution as a coarse stepwise probability density function (PDF), and (b) limit the numeric distribution to relatively short IRDs, using a general IRM frequency distribution (e.g., Zipf) to approximate the tail of long-duration ones.
This parsimonious parameterization creates standardized, reproducible trace profiles for consistent cross-system evaluation while allowing proprietary measurements of workload behavior to be distilled into a form which may be easier to make public, or shared among organizations.

\subsection{Contributions}
The contributions of 2DIO include:
\begin{enumerate}
    \item encoding complex workload characteristics using a succinct parameter triplet, requiring only a handful of numerical values,
    \item allowing the parameter space to be "swept" in experimentation, creating a continuum of complex LRU cache behaviors from exact reproduction of real-world patterns to hypothetical “what-if” scenarios,
    \item reproducing these behaviors while arbitrary scaling both footprint and length, allowing authentic and stressful system benchmarking at various scales. 
\end{enumerate}

The rest of the paper is organized as follows: Sec.~\ref{sec:background} covers background in workload models and real-world trace behaviors. Sec.~\ref{sec:2DIO} describes high level approach, introducing the two main algorithms. Sec.~\ref{sec:implementation} explains design and implementation details. Sec.~\ref{sec:evaluation} evaluates 2DIO's fidelity, configurability, scalability, and discusses limitations. Sec.~\ref{sec:related_works} reviews related research, and Sec.~\ref{sec:conclusion} concludes the paper.

\begin{figure*}[h!]
  \centering 
  \includegraphics[width=1\textwidth]{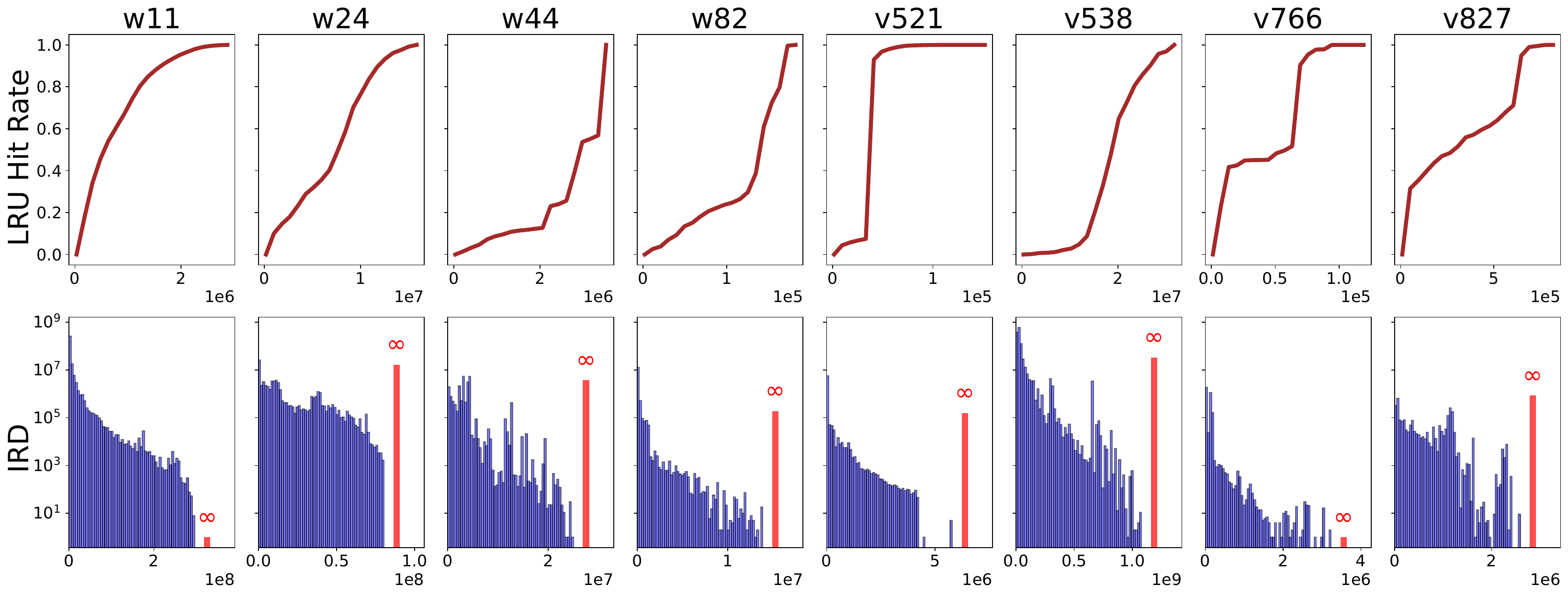}
  \caption{LRU HRCs and IRD histograms of real traces from {AliCloud} and {CloudPhysics}.}
  \label{fig:real}
  \vspace{-0.1in}
\end{figure*} 
\section{Background}
\label{sec:background}

\subsection{Workload models}
\label{sec:workload_models}
Cache algorithms have long been characterized by how they make use of \emph{recency} and \emph{frequency} of cached items, i.e. the time since an item's last occurrence and the rate at which the item has occurred in the past, respectively.
LRU uses only recency to make eviction decisions; least-frequently-used (LFU), on the other hand, uses only frequency. A broader spectrum of cache algorithms (e.g, First-In-First-Out (FIFO), CLOCK) are primarily recency-based, but respond to frequency as well.

In describing these models, we assume a footprint of $M$ distinct items (e.g., a block of 4096 bytes); a cache that can hold at most $C$ of them, never more; and a reference stream $r_1, r_2, \cdots$ of accesses to items in $\{0, 1, \cdots, M-1\}$. Following prior work~\cite{waldspurger_efficient_2015, waldspurger_cache_2017}, we simulate the reference stream as a discrete-time process, measuring time $t$ in distance (units of accesses). E.g. the distance between references $r_4$ and $r_1$ is $3$.

\noindent\textbf{Independent reference model.} IRM~\cite{aho1971principles, coffman_operating_1973} characterizes an access stream only by its item frequency; each item $i$ is assigned with weights $w_i, i \in \{0, 1, \cdots, M-1\}$, and $\sum_i w_i = 1$. Hence access to $i$ is chosen independently with probability $w_i$.

IRM workloads are widely used in microbenchmarks~\cite{fio, iometer, bonnie, Filebench}, typically in its simplified forms.
In the hot/cold model~\cite{rosenblum_design_1991},  $r_H M$ hot items are accessed at rate $\lambda_H$, and the remaining $(1-r_H ) M$ cold items are accessed at rate $1-\lambda_H$; in the Zipfian model, $\lambda_i$ corresponds to a Zipf distribution.

\noindent\textbf{Dependent reference models.}
Alternately one can ignore frequency\footnote{Frequency does not affect LRU performance, but affects FIFO and CLOCK, and may have greater effect on more modern replacement algorithms.} and characterize an access stream by the distribution of \emph{distances} between accesses to an individual item.

This approach has been widely used for \emph{cache modeling}, i.e. predicting cache performance from workload characteristics~\cite{mattson_evaluation_1970, eklov2010statstack, niu2012parda, wires_characterizing_2014, sen2013reuse, xiang2011linear, hu2015lama}.

Two measures have been used for this distance: \emph{stack distance}~\cite{mattson_evaluation_1970} (SD)  and \emph{inter-reference distance}~\cite{denning_working_1968,denning1972properties} (IRD).
Given two successive references $r_i$ and $r_j$ to the same item, the IRD refers to the distance between them in the reference stream (i.e. $j-i$), while the SD is the number of unique items referenced by the accesses separating them (i.e. $|\{r_{i+1}, r_{i+2}, \cdots, r_{j-1}\}|$).

\noindent\textbf{AET Approximation.} There is a direct correspondence between SD and LRU hit rate: if the SD between two references is $C$, then the second access will hit in any LRU cache of size $C$ or larger. 

This correspondence can be approximated to a correlation between IRD and LRU miss rate~\cite{hao_che_hierarchical_2002,hu_kinetic_2016}: if one knows the average time items stay in cache before eviction (AET), and the cumulative distribution function $F(t)$ of the request IRDs, one can approximate the cache size $C$ (i.e., SD). It follows that
\begin{equation*}
C = \int_{0}^{AET(C)} P(t)dt,
\end{equation*}
where $P(t) = 1-F(t)$ is the probability that an item is not reused before $t$, and $AET(C)$ denotes the average eviction time of items in this cache.

The miss rate at cache size $C$ is the probability that a reuse time is greater than $AET(C)$:
\begin{equation*}
 P_{miss}(C) = P(AET(C)).
\end{equation*}
This is known as Che's approximation~\cite{hao_che_hierarchical_2002}, or AET approximation~\cite{hu_kinetic_2016}. 

\subsection{Real Block Trace Characteristics}
\label{sec:real_traces}
IRM may well characterizes workloads seen by e.g. CDNs, where large numbers of independent sources are aggregated into a single request stream, ``drowning out'' correlations between references in a single stream.
Block storage workloads are different. Their references typically driven by one or a few applications, preserving inter-reference correlations caused by application behavior.

Table~\ref{tab:subset_chosen} lists several AliCloud (v521, v538, v766, v827) and CloudPhysics (w11, w24, w44, w82) traces. This subset was chosen for subsequent analysis (Secs.~\ref{sec:fidelity}, \ref{sec:scalability}) because they exhibit diverse cache behaviors. Length and footprint are measured after converting each trace to PARDA~\cite{niu2012parda} format\footnote{A sequence of 64-bit references without additional metadata, used for cache simulation.}. 
\begin{table}[h!]
  \centering
  \caption{Trace subset chosen for subsequent analysis.}
  \label{tab:subset_chosen}
  \begin{tabular}{lcc}
    \toprule
    Trace ID & Length (N) & Footprint (M)\\
    \midrule
    w11 & 296893045 & 2992519 \\
    w24 & 81762918 & 16487648 \\
    w44 & 25257814 & 3679382 \\
    w82 & 14198758 & 189785 \\
    v521 & 5974956 & 158018 \\
    v538 & 1204044775 & 33006370 \\
    v766 & 3335779 & 124146 \\
    v827 & 3198158 & 851527 \\
    \bottomrule
  \end{tabular}
\end{table}

Fig.~\ref{fig:real} depicts their IRD histograms alongside the corresponding LRU HRCs. 
HRCs are seen to be highly \emph{non-concave}, with steep sections (\emph{cliffs}) and flat regions (\emph{plateaus}). These correspond to the \emph{spikes} and \emph{holes} in the underlying IRD distribution.
Holes starting at zero may be due to the OS buffer cache \cite{willick_disk_1993}, which absorbs low-IRD accesses, while \emph{spikes} may be caused by scan-like behavior. 

A few of them display ``well-behaved'' IRM-like behaviors, e.g., {w11} has a decreasing IRD histogram which corresponds to a concave HRC.

All of them show a noticeable percentage of "one-hit wonders" which are referenced once and never re-accessed over the duration of the trace\footnote{CloudPhysics and AliCloud traces are $7$ and $30$ days long; any access beyond this time period is unlikely to hit in cache.}. In IRD measurements these accesses are recorded with an IRD of $\infty$.

\section{2DIO Framework}
\label{sec:2DIO}
The trace generation algorithm at the core of 2DIO can be considered a simple discrete-event simulation: each item repeatedly (a) generates one access to itself, (b) selects a sleep time $t$ from the IRD distribution, and (c) sleeps for time $t$. Since we focus on distances between accesses rather than actual time, the algorithms always generate exactly one access for every position in the trace produced.

% We reserve a fraction of the access footprint, referred to as the singleton pool, for "one-hit wonders". 

% The IRD distribution and footprint must align for scalability. For a footprint of $M$ items (excluding one-hit wonders), the mean IRD must also be ${M}$.

\subsection{2DIO Generation from IRD}
\label{sec:gen_from_ird}
\RestyleAlgo{ruled}
\begin{algorithm}[h!]
  \SetAlgoLined
  \KwIn{$f$, $M$, $N$}
  \KwOut{$\pi_s$}
  \BlankLine
  $\text{Heap} \gets \emptyset$ \\
  $a \gets 0$ \\
  % \\\tcc{Assign an address to each drawn IRD, and push all pairs to the Heap}
  \While{$\text{Heap.size} < M$}{
    $t \overset{\text{i.i.d.}}{\sim} f$\\
    \If{$t \neq \infty$}{
      $\text{Heap.insert}(<t,\  a>)$ \\
      $a \gets a + 1$
    }
  }
 
  $\pi_s \gets \emptyset$ \\
  \For{$j = 0 \ldots N-1$}{ 
    $t \overset{\text{i.i.d.}}{\sim} f$ \\
    \If{$t = \infty$}{
    % \tcc{Assign a new address that is not on the Heap}
      $\pi_s.\text{append}(a)$ \\
      $a \gets a + 1$
    }
    \Else{
        % \tcc{Append the lowest address to $\pi_s$}
      $<t_0,\  a_0> \gets \text{Heap.pop}()$ \\
      $\pi_s.\text{append}(a_0)$ \\
      % \tcc{Update its time track with $t_0 + t$}
      $\text{Heap.replace}(<t_0 + t,\  a_0>)$
    }
  }
  \KwRet{$\pi_s$}
  \caption{\textbf{Gen-from-IRD}: generate a trace from a given inter-reference distance (IRD) distribution $f$.} 
  \label{alg:gen_from_ird}
\end{algorithm}
Algorithm~\ref{alg:gen_from_ird} presents an effective way of generating a trace from a given IRD distribution.

\noindent \textbf{Input.} The algorithm takes the following input parameters: 
\begin{enumerate}
  \item[(1)] $f$ -- an IRD distribution
  \item[(2)] $M$ -- footprint 
  \item[(3)] $N$ -- trace length
\end{enumerate}

\noindent \textbf{Output:} Synthetic trace $\pi_s$. 

\noindent \textbf{Initialization:}
Draw $M$ sleep times $t$ from $f$. For each non-infinite $t$, assign a unique address $a$ to it and add the pair to the priority queue (heap).

\noindent \textbf{Trace generation.} To generate a single reference, draw a $t$ from the IRD distribution. If $t=\infty$ we draw an item from the singleton pool (addresses beyond $M$), otherwise we take the pair $\langle t_0,a_0\rangle$ from the bottom of the priority queue, generate an access $a_0$ to the trace, and update $a_0$'s sleep time to $t_0+t$ in the priority queue. The algorithm continues until the trace is of length $N$.

% We describe a high-difelity trace regeneration process utilizing Algorithm~\ref{alg:gen_from_ird}: given an empirical measurements of an IRD distribution, e.g. a fine-grained histogram or random selection from the full set of measured IRDs, the algorithm generates synthetic traces $\pi_{s}$ which closely mimic the behavior of the original trace $\pi_o$.

If one is only interested in shaping the LRU or other purely recency-based HRCs, then this algorithm is sufficient, for IRDs alone dictate the performance of these cache policies~\cite{denning1972properties,hao_che_hierarchical_2002}.
Policies such as FIFO and CLOCK, however, also respond to frequency rather than recency alone. We thus describe an extension to Gen-from-IRD which induces tunable popularity skew.

\subsection{2DIO Generation from IRD + IRM}
\RestyleAlgo{ruled}
\begin{algorithm}[h!]
  \SetAlgoLined
  \KwIn{$P_{IRM}$, $g$, $f$, $M$, $N$}
  \KwOut{$\pi_s$}
  \BlankLine
  $\text{Heap} \gets \emptyset$ \\
  $a \gets 0$ \\
  % \tcc{Assign an address to each drawn IRD, and push all pairs to the heap}
  \While{\text{Heap.size} $< M$}{
    $t \overset{\text{i.i.d.}}{\sim} f$ \\
    \If{$t \neq \infty$}{
      $\text{Heap.insert}(<t, a>)$ \\
      $a \gets a + 1$
    }
  }
  % \tcc{Create trace}
  $\pi_s \gets \emptyset$ \\
  \For{$j = 0 \ldots N-1$}{
    $Random \overset{\text{i.i.d.}}{\sim} \text{Uniform}(0, 1)$ \\
    \If{$\text{Random} < P_{IRM}$}{
      $\text{addr}  \overset{\text{i.i.d.}}{\sim} g$ \\
      %  \tcc{Sample a reference address from g}
      $\pi_s.\text{append}(\text{addr})$
    }
    \Else{
      $t \overset{\text{i.i.d.}}{\sim} f$ \\
      \If{$t = \infty$}{
        % \tcc{Assign a new address that is not on the heap}
        $\pi_s.\text{append}(a)$ \\
        $a \gets a + 1$
      }
      \Else{
        % \tcc{Append the lowest address to $\pi_s$}
        $<t_0, a_0> \gets \text{Heap.pop}()$ \\
        $\pi_s.\text{append}(a_0)$ \\
        % \tcc{update time track with $t_0 + t$}
        $\text{Heap.replace}(<t_0 + t, a_0>)$
      }
    }
  }
  \KwRet{$\pi_s$}
  \caption{\textbf{Gen-from-2D}: generate a trace from the IRD distribution $f$ with probability $1 - P_{IRM}$, and from the item-frequency distribution $g$ with probability $P_{IRM}$.}
  \label{alg:gen_from_both}
\end{algorithm}

Gen-from-IRD relies heavily on the accuracy of the input IRD distribution, often requiring it to be long enough to manifest a long tail. IRDs beyond the eviction time do not affect the \emph{non-concave} HRC features and thus can instead be approximated by a succinct IRM input (e.g., Zipf (1.2)), while explicitly enforcing the frequency distribution.

\begin{figure}[ht!]
  \centering
  \includegraphics[width=0.9\columnwidth]{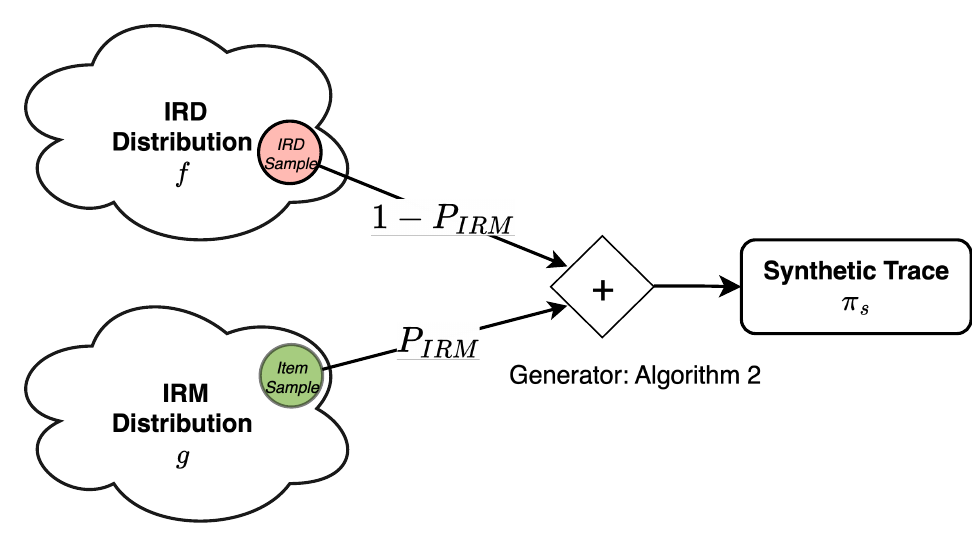}
  \caption{IRD + IRM trace generation uses Algorithm~\ref{alg:gen_from_both}: references are drawn with probability $P_{IRM}$ from an IRM process, and $1-P_{IRM}$ from an IRD renewal process.}
  \label{fig:2d_trace_gen}
  \vspace{-0.1in}
\end{figure}

We introduce Algorithm~\ref{alg:gen_from_both}: Gen-from-2D. It is essentially derived from Gen-from-IRD, but merged with an IRM arrival process. In this process, item frequency follows distribution $g$, and each generation of such items is triggered with probability $P_{IRM}$. The process is visualized in Fig.~\ref{fig:2d_trace_gen}. 
Merging a short-interval IRD distribution $f$ with a long-interval IRM tail $g$ lets the generator tune both dimensions, shaping the behavior of diverse cache policies.

\noindent \textbf{Input:} The algorithm takes $5$ inputs:
\begin{enumerate}
    \item[(1)] $P_{IRM}$ -- fraction of independent references
    \item [(2)]$g$ -- an item frequency distribution
    \item[(3)] $f$ -- IRDs represented as a piece-wise quantization
    \item[(4)] $M$ -- footprint
    \item[(5)] $N$ -- trace length
 \end{enumerate}

\noindent \textbf{Output:} Synthetic trace $\pi_s$.

\noindent \textbf{Initialization:} see Gen-from-IRD.

\noindent \textbf{Trace generation:} The algorithm is similar to Gen-from-IRD. However, at each iteration, with probability $P_{IRM}$, it chooses an item from $g$.

\subsection{Parameters Customization}
\label{sec:type_a}

Following Algorithm~\ref{alg:gen_from_both}, 2DIO takes in $5$ inputs: $P_{IRM}$, $g$, $f$, $M$, and $N$.
The first three we define as the \emph{trace profile}, represented as a triplet $\theta = \langle P_{IRM}, g, f \rangle$. We refer to items generated by $f$ as \emph{dependent arrivals}, and those by $g$ as \emph{independent arrivals}.
$M$ and $N$ are {scale parameters} which do not affect the (normalized) cache behavior.

The goal of 2DIO is to use a compact $\theta$ to create synthetic traces from scratch; it can also reproduce real traces by calibrating $\theta$ to match similar behavior.
For both cases, 2DIO aims to approximate the target cache behavior, rather than reproducing it exactly, and that a given accuracy target may be met by more than one choice of $\theta$. 

Essentially, there is a trade-off between accuracy and succinctness: reproducing a real trace may require a long or empirically derived $f$ (as in Fig.~\ref{fig:simple_HRC}), whereas generating new behavior requires only a succinct $f$ (e.g., fewer than 10 values with \emph{spikes} at target percentiles), offering high tractability.

\subsubsection{Configuring the IRD Distribution.}
\label{sec:fgen}
$f$ specifies an IRD distribution from which the algorithm draws the sleep time for each item.
A typical scenario is setting up a trace that has a cache benefit when the cache size is around e.g., 5\% of the dataset. As proven by van den Berg and Towsley (1993)~\cite{van1993properties}, this cannot be achieved by adjusting the frequency distribution; it requires shaping the underlying IRD distribution, e.g., inducing \emph{spikes/holes} at specific values. 

We note that this can be implemented with the AET approximation formulas~\cite{hao_che_hierarchical_2002,hu_kinetic_2016} presented in Sec.~\ref{sec:workload_models}:
\begin{equation}
  C = SD(\mtc)= \int_{0}^{\mtc} P(t)dt,
  \label{eq:che}
  \end{equation}
Since Eq.~\eqref{eq:che} is bijective, each mean eviction time \tc on the IRD domain uniquely maps to a cache size $C$ and vice versa.

The corresponding hit rate is given by:
\begin{equation}
  P_{hit}(C) = 1 - P(\mtc),
 \label{eq:mr}
 \end{equation}
whereby each \tc also yields a unique hit rate for the stack distance $C$ it approximates.
\begin{figure}[h!]
  \begin{center}
  \subfigure[\emph{plateau} in HRC at $C_i$ to $C_j$ $\leftrightarrow$ \emph{hole} in $f$ at $AET(C_i)$ to $AET(C_j)$]
  {
  \includegraphics[width=1\columnwidth, height=0.4\columnwidth]{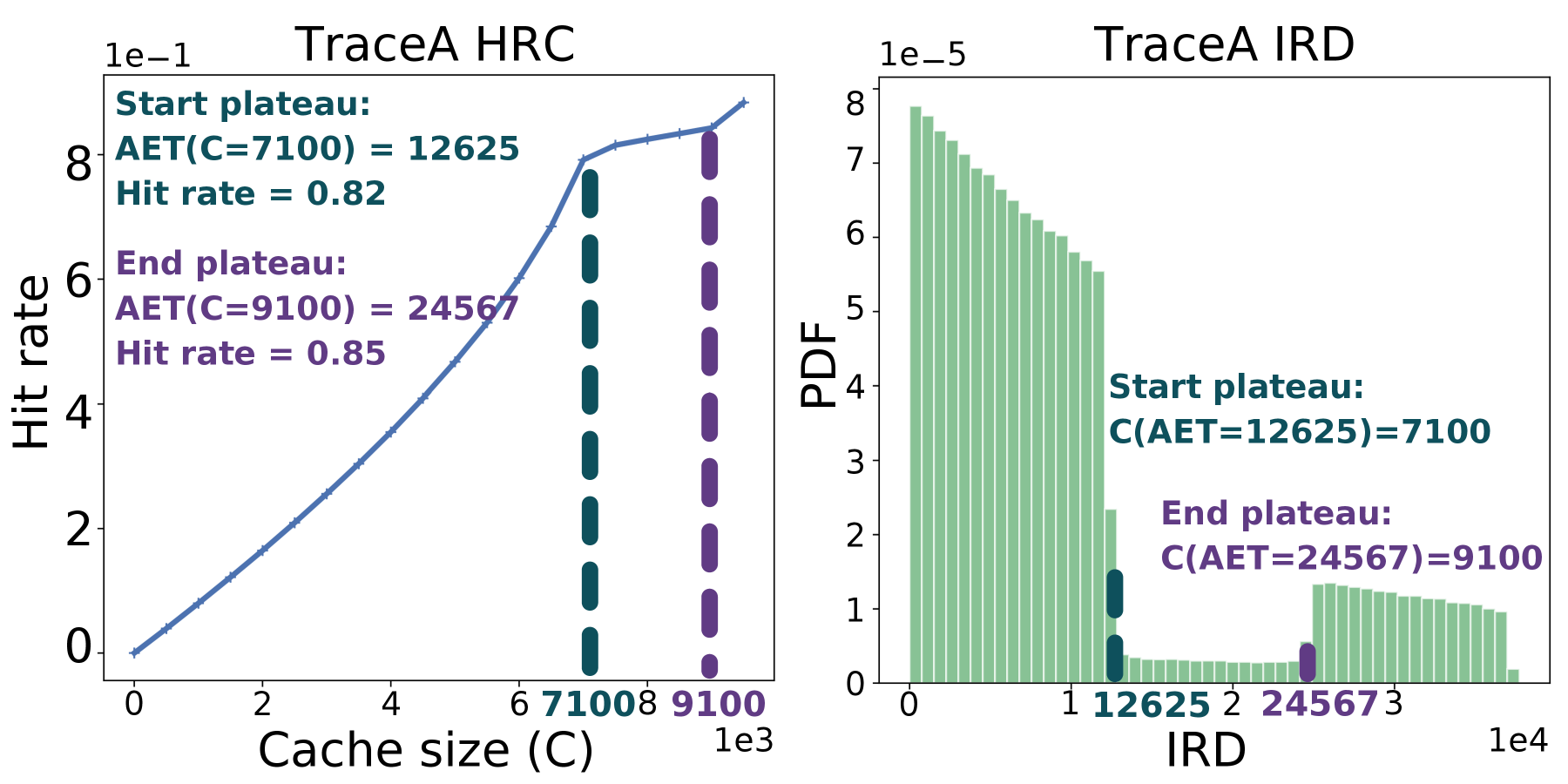}
  \label{fig:ta_2_plot}
  }
  \subfigure[\emph{cliff} in HRC at $C_i$ to $C_j$ $\leftrightarrow$ \emph{spike} in $f$ at $AET(C_i)$ to $AET(C_j)$]
 { 
  \includegraphics[width=1\columnwidth, height=0.4\columnwidth]{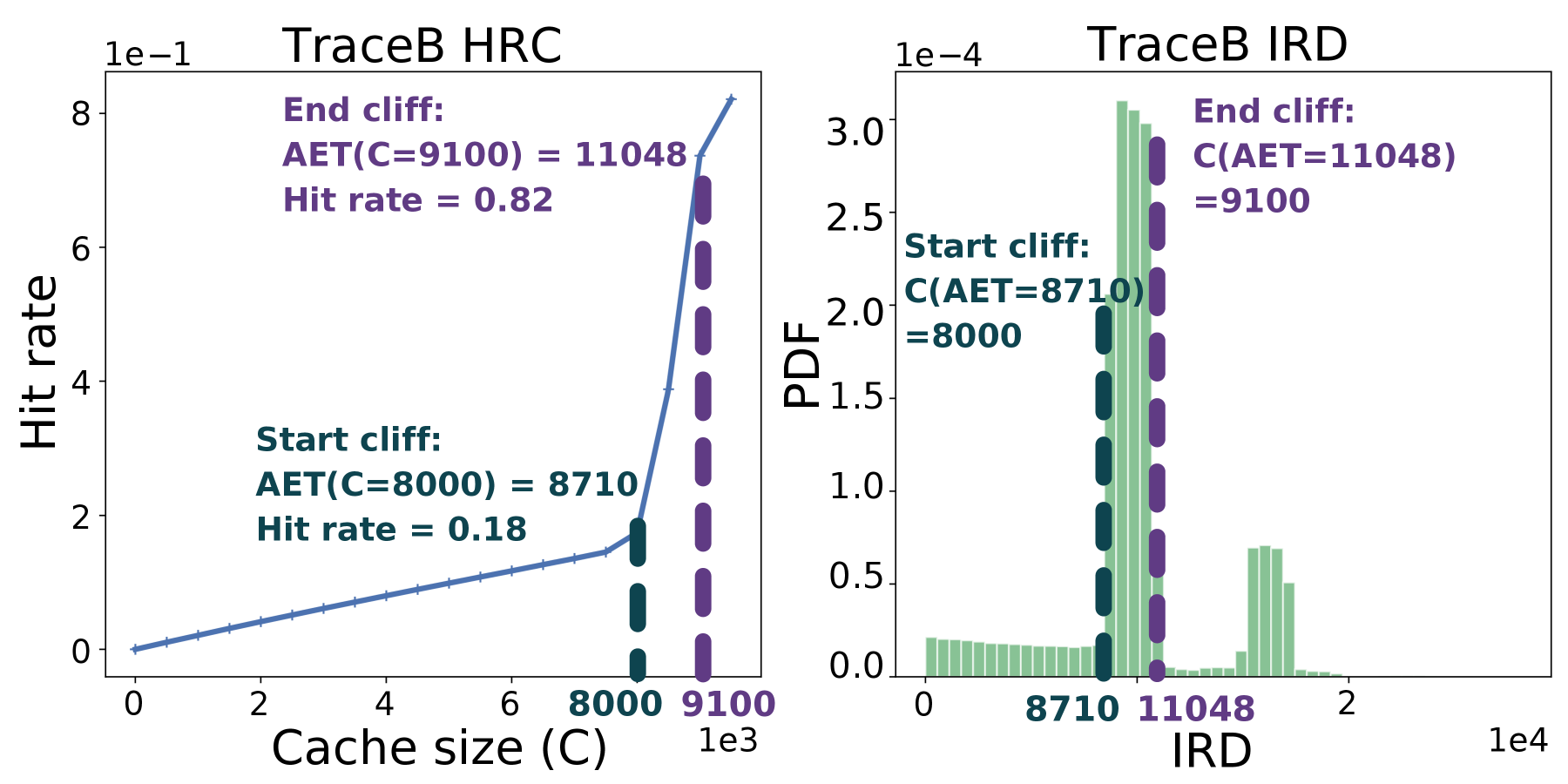}
  \label{fig:tb_2_plot}
 }
  \end{center}
  \caption{correspondences between HRC \emph{plateaus} \& \emph{cliffs} and IRD \emph{holes} \& \emph{spikes}. Example synthetic traces {Trace A} and {Trace B}: LRU HRC (left), and corresponding IRD distributions (right).}
  \label{fig:2_plot}
  \vspace{-0.1in}
\end{figure}

Fig.~\ref{fig:2_plot} visualizes such correspondence between a \emph{hole} in the {TraceA} IRDs and a \emph{plateau} in its HRC, as well as a \emph{spike} in {TraceB} IRDs and a \emph{cliff} in its HRC. 
To utilize this correspondence, 2DIO provides a model-accurate interface \textbf{fgen}.
Let the IRD distribution $f(X = i)$ be a PMF with finite support $i \in \{1, 2, \cdots, k\}$.
Let \textbf{fgen}$(k, \mathcal{I}, \epsilon)$ be a function which generates $f$, assigning higher probability (\emph{spikes}) to elements in the set $\mathcal{I}$ and lower probability (\emph{holes}) to elements in its complement set $\overline{\mathcal{I}}$. The total probability mass of the \emph{holes} sums to $\epsilon$, resulting in the following distribution:
\begin{equation}
  f(i) =
  \left\{
    \begin{array}{ll}
      \frac{1 - \epsilon}{\lvert \mathcal{I} \rvert}, \quad i \in \mathcal{I} \\
      \frac{\epsilon}{k-\lvert \mathcal{I} \rvert}, \quad i \in \overline{\mathcal{I}}
    \end{array}
  \right.
  \label{eq:fgen}
\end{equation}
2DIO then fit $f$ over an auto-tuned IRD sample space (see Sec.~\ref{sec:ird_sampler}) $\mathcal{S} := \{1, 2, \cdots, T_{max}\}$, result in a distribution with \emph{spikes} at IRD intervals: 
$$\{[i \times \frac{T_{max}}{k}, \ (i+1) \times \frac{T_{max}}{k}], \quad \forall i \in \mathcal{I}\}$$ 
and \emph{holes} elsewhere, which manifest as \emph{cliffs} in HRC at cache size intervals:
$$
\{[SD(i \times \frac{T_{max}}{k}), \ SD((i+1) \times \frac{T_{max}}{k})], \quad \forall i \in \mathcal{I}\}
$$
and \emph{plateaus} elsewhere in the $C$ domain, following Eq.~\eqref{eq:che}.

\subsubsection{Selecting the IRM Distribution.}
$g$ specifies an item frequency distribution from which the algorithm selects items directly and adds to the trace. The choice of $g$ determines the item-frequency distribution (i.e., popularity). $P_{IRM}$, in turn, decides the fraction of these arrivals in the generated trace.

We note that $f$ is a finite distribution, generating IRDs between $0$ and a maximum value $T_{max}$. Fitting $P_{IRM}$ and $g$ involves examining the IRD distribution for values beyond $T_{max}$. The final IRD distribution is a merge of IRDs of dependent arrivals and independent arrivals, with the latter contributing to the tail.

% For highly detailed and delicate configuration, we manually select $P_{IRM}$ and a $g$ to fit the IRD tail above $T_{max}$, followed by fitting $f$. This process is iterated by adjusting $P_{IRM}$ and repeating until the desired accuracy is achieved. Our toolkit provides an interactive visualization tool to assist with this process.
\subsubsection{Exploring Parameter Space for Desired Trace Behavior.}
\label{sec:sweep}
In practice, users don't need to understand the underlying fitting model. 2DIO provides an interactive visualization tool to assist parameter tuning. Users can (1) select a cache algorithm from the built-in {cachesim} library, (2) start with intuitive (or default) inputs, then (3) drag a slider or type in a number to change any single parameter and immediately observe the HRC recomputed on the fly.

While simulating the HRC on the fly can be resource-intensive, using a small trace footprint $M$ and length $N$ (e.g., 100, 10000, respectively) during this process minimizes overhead. Once tuned, parameters are reusable at various scales without compromising fidelity (see Sec.~\ref{sec:scalability}).

%-------------------------------------------------------------------------------
\section{Implementation}
\label{sec:implementation}
\begin{figure*}[t]
  \begin{center}
  \subfigure[TraceA: $P_{IRM}=0.1$, $g = \text{Zipf}(1.2)$, $f$ fitted empirically to a simple 3-class IRD distribution]
  {
  \includegraphics[width=2\columnwidth]{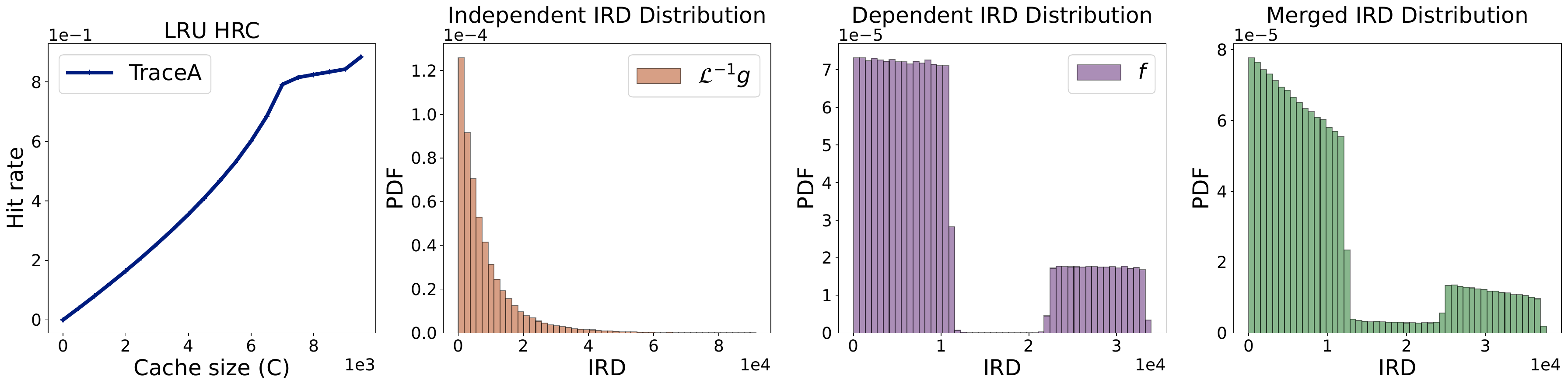}
  \label{fig:ta_4_plot}
  }
  \subfigure[TraceB: $P_{IRM}=0.1$, $g = \text{Pareto}(2.5, 1)$, $f$ fitted empirically to a 15-class IRD distribution]
 { 
  \includegraphics[width=2\columnwidth]{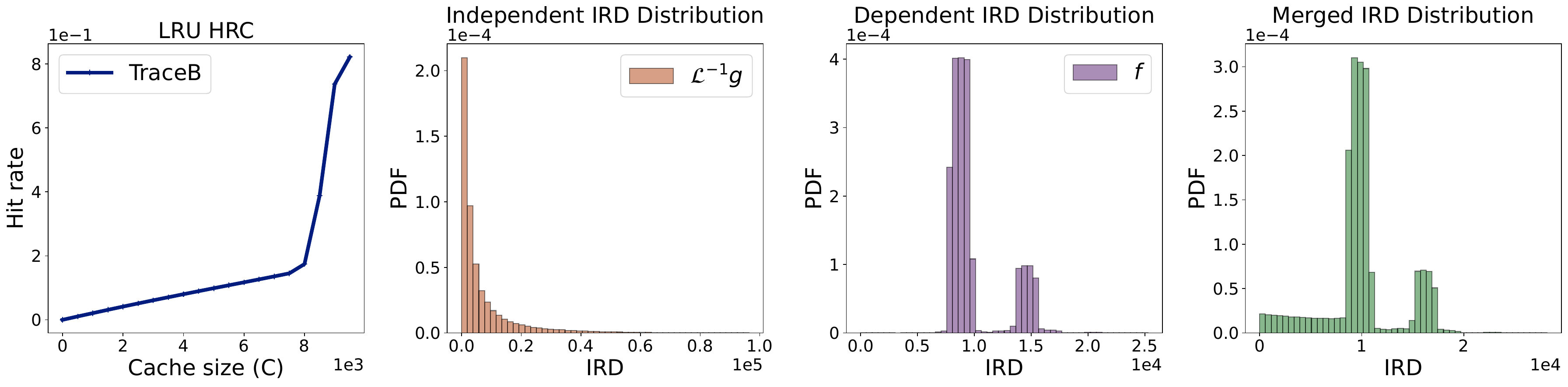}
  \label{fig:tb_4_plot}
 }
  \end{center}
  \caption{HRCs for TraceA and TraceB, with separate visualizations of IRD histograms for dependent, independent, and merged arrivals. $\mathcal{L}^{-1}{g}$ denotes the inverse Laplace transform of $g$, mapping its frequency-domain representation to the IRD domain.}
\label{fig:4_plot}
\end{figure*}
2DIO package\footnote{Available at \url{https://github.com/Effygal/trace-gen}; artifact also on \url{https://zenodo.org/records/17202588}.} includes a CLI tool \texttt{trace-gen} written in C++ for trace generation, and a Python library for cache simulation and analysis. \texttt{trace-gen} takes as input the footprint $M$, trace length $N$, and \emph{trace profile} $\theta = \langle P_{\text{IRM}},\ g,\ f \rangle$.

Since we focus on block-level HRCs, all access units are assumed uniform, which is typical in storage cache evaluation.

This section details the implementation of the IRD and IRM samplers, introduces several default \emph{trace profiles}, and provides example commands for generating traces with target cache behaviors.

\subsection{Sampling Dependent Arrivals}
\label{sec:ird_sampler}
\texttt{trace-gen} accepts $f$ input through the \textbf{fgen} interface. We have preconfigured six default \emph{trace profiles} for the toolkit. These generalize several real-world IRD patterns, labeled $\hat{f}_a$ through $\hat{f}_g$ (see Appendix~\ref{sec:default_vis} Table~\ref{tab:parameters}). Their corresponding HRCs and IRD distributions are visualized in Fig.~\ref{fig:canonical_mrc_ird} of the same appendix.

\noindent\textbf{Auto-tuned IRD sample space.}
 $f$ is fitted to an auto-tuned $k$-binned sample space $\mathcal{S}$. The IRD sampler in Algorithm~\ref{alg:gen_from_both} selects a bin $i$ with probability $f(i)$ and samples an IRD uniformly from that bin.

To ensure $\mathcal{S}$ matches the empirically measured IRD distribution from target trace $\pi_s$, $T_{max}$ is auto-tuned based on $M$ such that the mean of drawn IRD samples equals $M$. It follows that $$M = \sum_{i=1}^{k} b_i \cdot f(i),$$ where $b_i$ is the midpoint of bin $i$: $$b_i = \frac{(2i - 1)}{2} \times \frac{T_{max}}{k}.$$ Hence $T_{max}$ is solved by: $$T_{max} = \frac{2Mk}{\sum_{i=1}^{k} (2i - 1) \cdot f(i)}.$$

\subsection{Sampling Independent Arrivals}
The item-frequency distribution $g$ for independent arrivals can follow Zipf, Pareto, Normal, or Uniform distributions over a sample space $\mathcal{U}$. \texttt{trace-gen} accepts the IRM type as a string input, defaulting to "zipf" with $\alpha = 1.2$ if unspecified.
For each specified IRM type, the associated parameters can be individually configured, generating a PMF $g$ as shown in Table~\ref{tab:irm_pmf}. The IRM sampler selects an item $i$ from $\mathcal{U}$ with probability $g(i)$ following the corresponding PMF.
% \vspace{-0.3cm}
\renewcommand{\arraystretch}{1.2}
\begin{table}[ht]
  \caption{\texttt{trace-gen} supported IRM types and corresponding PMF expressions.}
  \vspace{0.5em}
    \centering
    \resizebox{\columnwidth}{!}{%
    \begin{tabular}{ccc}
      \toprule
        \textbf{IRM Type} & \textbf{Configurable} & \textbf{PMF} \\
        \midrule
        Zipfian    & $\alpha$              & $g(i) = \bigl(\tfrac{1}{i}\bigr)^{\alpha}$ \\ 
        Pareto     & $\alpha,\,x_m$        & $g(i) = \bigl(\tfrac{x_m}{i}\bigr)^{\alpha}$ \\ 
        Normal     & $\mu,\,\sigma$        & $g(i) = \frac{1}{\sigma\sqrt{2\pi}} \exp\ \bigl(-\tfrac{(i-\mu)^2}{2\sigma^2}\bigr)$ \\ 
        Uniform    & $a=0,\,b=M-1$         & $g(i) = \frac{1}{M}$ \\ 
        Empirical  & counts $n_i$          & $g(i) = \frac{n_i}{\sum_{j}n_j}$ \\ \bottomrule
    \end{tabular}%
    }
    \label{tab:irm_pmf}
\end{table}

\subsection{2D Generation Command}
To generate {TraceA} with $f = \hat{f}_b$ and default $g = \text{Zipf}(1.2)$, use:
\begin{lstlisting}
trace-gen -m 10000 -n 1000000 -f b -p 0.1
\end{lstlisting}
The resulting HRC and IRDs for independent, dependent, and merged arrivals are visualized in Fig.~\ref{fig:ta_4_plot}.
To generate {TraceB} with an explicitly specified $f=\text{Pareto}(2.5, 1)$, use:
\begin{lstlisting}
  trace-gen -m 10000 -n 1000000 -f fgen:15:0.01:1,3,5,9 -g pareto:2.5,1 -p 0.1
\end{lstlisting}
See Fig.~\ref{fig:tb_4_plot} for the same visualization.
 In both traces, 90\% of arrivals are sampled from $f$ and only 10\% from $g$, yielding minor independent influence. As a result, both traces exhibit highly \emph{non-concave} HRCs.

% \begin{figure*}
%     \centering
%     \includegraphics[width=2.0\columnwidth]{./figures/interactive.png}
%     \caption{Interactive visualization tool for LRU HRC produced by 2DIO trace-gen parameters.}
%     \label{fig:interactive}
%   \end{figure*}
One can ingest an I/O size distribution to vary request sizes if specified. E.g., \texttt{--sizedist 1,1,1:1,3,4} means equal chances of 1-, 3-, or 4-block requests. However, doing so may affect the carefully crafted IRD distribution. See Sec.~\ref{sec:limitations} of this limitation.

% \subsection{{Type (b) TraceSynthesizer}}
% \texttt{TraceSynthesizer}  is used to generate type (\textbf{b}) synthetic traces based on a given real trace. The API provides two methods for synthetic trace-reconstruction: inter-reference distance-based (IRD) reconstruction and item frequency-based (IRM) reconstruction.

% The method is to compute the IRD descriptor $\langle P_{single}, f \rangle$ introduced in Section~\ref{sec:real_synthesis} of the input original real trace, and reconstruct synthetic traces  of length $n$ with the same IRD descriptor, assume the original trace $\pi_o$ is pre-processed in PARDA format,
% \begin{verbatim}
% regenerator = tg.TraceSynthesizer(pi_o)
% \end{verbatim}
% IRD-based synthetic trace reconstruction:
% \begin{verbatim}
% pi_s = regenerator.trace_gen(n=100000)
% \end{verbatim}
% % IRM-based reconstruction:
% % \begin{verbatim}
% % trc_irm_recon = r1.generate_irm_trace(n=100000)
% % \end{verbatim}

\section{Evaluation}
\label{sec:evaluation}
This section evaluates 2DIO in terms of fidelity, configurability, and scalability. 
Evaluations involving real traces are conducted on the same subset described in Sec.~\ref{sec:real_traces}.
All cache simulations are performed using our built-in cachesim library. 

Because the LRU hit ratio is entirely determined by recency \cite{mattson_evaluation_1970, denning_working_1968}, we use LRU HRCs to assess how accurately our method produces/reproduces recency which has not been achieved by existing generators.

All simulations are run on Ubuntu 24.04 LTS with AMD Ryzen 5 7600 6-Core Processor and 62 GB RAM. GAN hyperparameter searching and training jobs are run on an HPC node with 1 NVIDIA Tesla V100-SXM2 GPU, 2 CPU cores, and 8 GB of allocated memory.

\subsection{Reproducing Real Trace with Succinct Parameters}
\label{sec:fidelity}
\begin{figure*}[h]
    % \centering  
    \includegraphics[width=2\columnwidth]{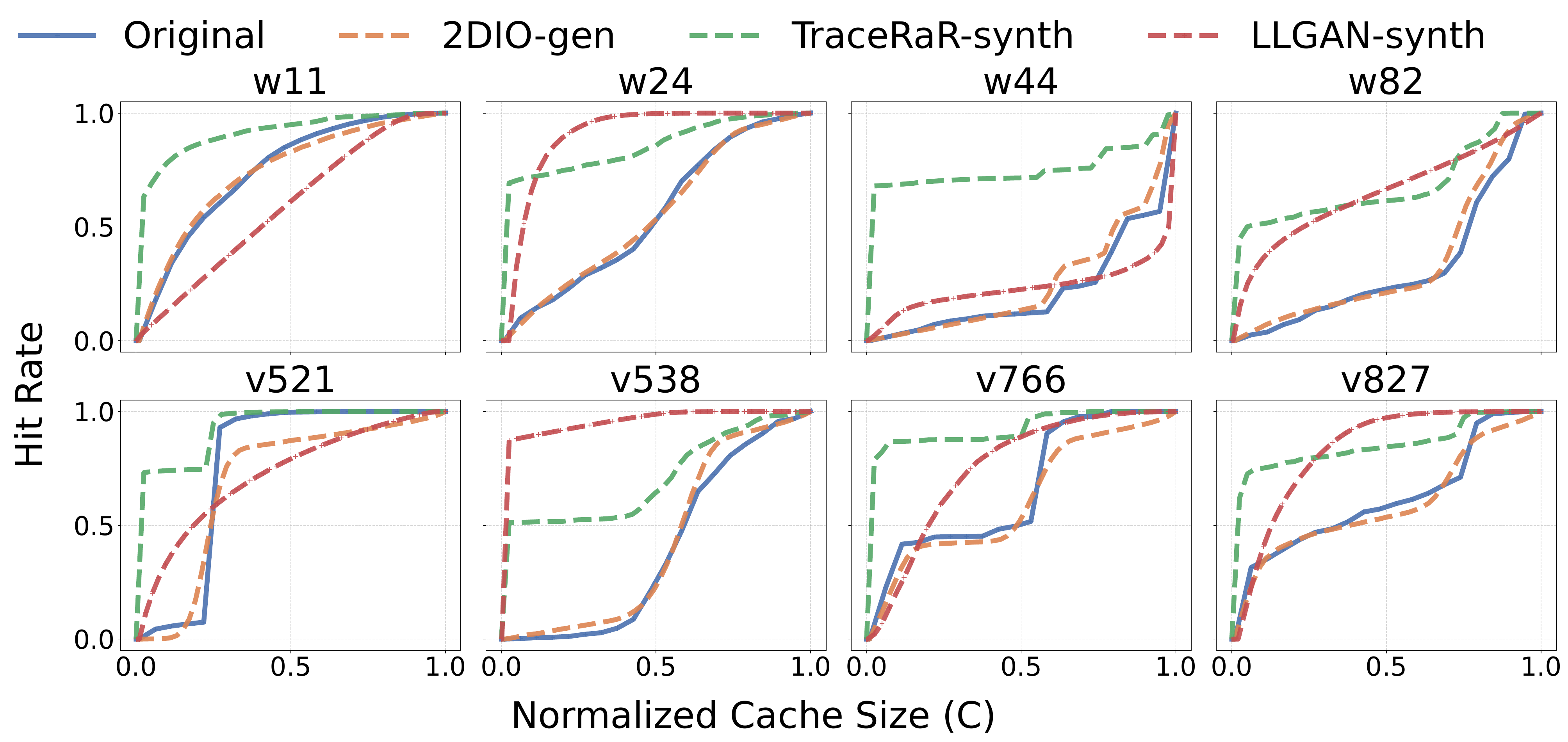 }
    \caption{2DIO reproducing real-world trace HRCs compared to TraceRaR and LLGAN synthesized traces.}
    \label{fig:real_vs_param}
\end{figure*}
Fidelity is assessed through HRC accuracy at block granularity. To test this we calibrate the profile $\theta$ for each trace, aiming to regenerate synthetic ones that produce similar HRCs---see Table~\ref{tab:theta_values}, all of which are expressed with less than $10$ numeric values. We then regenerate each on a small footprint $M=100$ and length $N=10$k except one\footnote{{w44} may require $M\geq10$k and $N\geq 1$m for a high-resolution HRC-reproduction.}, effectively reducing cache simulation costs.
\begin{table}[h]
    \centering
    \caption{Parsimonious trace profiles for "counterfeiting" each real-world trace. \textbf{fgen} follows Eq.~\eqref{eq:fgen}.}
    \vspace{0.5em}
    \resizebox{\columnwidth}{!}{%
    \begin{tabular}{lccc}
    \toprule
    \textbf{Trace ID} & $P_{IRM}$ & $g$ & $f$ \\
    \midrule
    \texttt{w11}       & 1.0 & $\mathrm{Zipf}(1.3)$ & $\mathrm{None}$ \\
    \texttt{w24}       & 0.45 & $\mathrm{Zipf}(1.2)$ & $\mathbf{fgen}(30, [1,2], 5e-3)$ \\
    \texttt{w44}       & 0.0 & $\mathrm{None}$ & $\mathbf{fgen}(30, [9, 13, 17, 19], 2.5e-2)$ \\
    \texttt{w82}       & 0.2 & $\mathrm{Zipf}(1.2)$ & $\mathbf{fgen}(100, [12, 13, 19], 1e-3)$ \\
    \texttt{v521} & 0.0 & $\mathrm{None}$ & $\mathbf{fgen}(100,[2],2e-3)$ \\
    \texttt{v538} & 0.1 & $\mathrm{Zipf}(1.2)$ & $\mathbf{fgen}(40,[3, 4],5e-3)$ \\
    \texttt{v766} & 0.0 & $\mathrm{None}$ & $\mathbf{fgen}(40, [0,5],5.7e-3)$ \\
    \texttt{v827} & 0.2 & $\mathrm{Zipf}(1.2)$ & $\mathbf{fgen}(60, [0, 13],5e-3)$ \\
    \bottomrule
    \end{tabular}%
    }
    \label{tab:theta_values}
\end{table}

Fig.~\ref{fig:real_vs_param} compares the HRCs of each 2DIO-generated trace (orange dashed) with their corresponding originals (blue solid). All relative \emph{cliff} and \emph{plateau} positions are preserved at cache sizes normalized to footprint $M$, with only minor deviations.

\subsubsection*{Comparison to Prior Work.}
\label{sec:alternative_tools}
\begin{table*}[h]
    \centering
    \caption{Optuna-optimized hyperparameters for LLGAN synthesis with resulted D-loss and MMD$^2$.}
    \vspace{0.5em}
    \resizebox{\textwidth}{!}{
    \begin{tabular}{c|c|c|c|c|c|c|c|c|c|c|c|c}
    \hline
     \textbf{Trace} & \textbf{Rand Seed} & \textbf{Hidden dim} & \textbf{Latent dim}  & \textbf{Batch size} & \textbf{Sequence len} & \textbf{Epochs} & \textbf{G-rate} & \textbf{D-rate} & \textbf{G-updates} & \textbf{D-updates} & \textbf{D-loss} & \textbf{MMD$^2$} \\
    \hline
    w11 & 77 & 126 & 10 & 64  & 12 & 30 & 0.000132 & 0.000454 & 1 & 2 & 0.5411 & 0.005128  \\
     w24 & 77 & 108 & 10 & 128 & 12 & 30 & 0.000090 & 0.000397 & 2 & 3 & 0.5308 & 0.071148 \\
     w82 & 42 & 121 & 10 & 128 & 12 & 30 & 0.000075 & 0.000108 & 1 & 2 & 0.4856 & 0.047067 \\
     w44 & 42 & 100 & 16 & 128 & 12 & 30 & 0.000248 & 0.000139 & 3 & 2 & 0.5600 & 0.060406 \\
     v521 & 77 & 101 & 10 & 64  & 12 & 30 & 0.000099 & 0.000157 & 1 & 3 & 0.5079 & 0.118840 \\
     v538 & 42 & 102 & 10 & 64  & 12 & 30 & 0.000415 & 0.000281 & 1 & 3 & 0.5510 & 0.016633 \\
     v827 & 77 & 108 & 10 & 128 & 12 & 30 & 0.000090 & 0.000397 & 2 & 3 & 0.5099 & 0.080228 \\
     v766 & 77 & 111 & 10 & 64  & 12 & 30 & 0.000254 & 0.000141 & 1 & 3 & 0.4818 & 0.013860 \\
    % 77  & w44 (Trial 2) & 103 & 16 & 128 & 12 & 30 & 0.000191 & 0.000179 & 1 & 1 & 0.075957 & 0.6100 \\
    \hline
    \end{tabular}
    }
    \label{tab:llgan_hyperparameters}
\end{table*}

TraceRaR~\cite{li2017tracerar} and generative adversarial networks (GANs)~\cite{zhang2024accurate} represent state-of-the-art trace synthesize methods that are optimized for I/O replay-accuracy. Here we evaluate whether such optimizations also preserve HRC fidelity.

\textbf{TraceRaR.}
TraceRaR~\cite{li2017tracerar} extends a given trace while preserving request rates, read/write ratios, and offset characteristics for effective scale-up replaying.

We used TraceRaR to extend each trace to twice its original length. In the synthesized output, the first half is identical to the original trace and therefore yields the same HRC. The second half, however, appears to be IRM; appending this segment to the real trace disrupts its recency. Consequently, the green dashed HRCs in Fig.~\ref{fig:real_vs_param} diverge markedly from the original.

\textbf{LLGAN.} LLGAN~\cite{zhang2024accurate} utilizes a one-layer long short-term memory (LSTM) architecture for both the generator (G) and discriminator (D), and have both optimized against cross-entropy losses. 

Since the original implementation is not publicly available and hyperparameter tuning is workload-specific, we reproduced the paper’s proposed design\footnote{Available at \url{https://github.com/Effygal/gan-io}}. 
Each trace is trained on two dimensions: \texttt{[LBA, Length]}\footnote{Only logical block addresses (LBAs) and I/O sizes (lengths) are relevant to cache performance.}. We run several rounds of 50-trial hyperparameter searches with Optuna\cite{akiba2019optuna}, looking for combinations where the D-loss settles around $0.5$ and the G-loss is minimized, which often indicates a well-trained GAN.
Table~\ref{tab:llgan_hyperparameters} summarizes the optimized hyperparameters for each trace, including random seeds used for reproducibility.

\textbf{Sanity check.} Before drawing any conclusion on HRC accuracy, we first validate the LLGAN-synthesized traces against the evaluation metric used in the original paper---the maximum mean discrepancy\footnote{Computed with an RBF kernel and median bandwidth $\sigma$.} (MMD$^2$), computed as the joint distributional difference over LBAs and lengths. The last two columns of Table~\ref{tab:llgan_hyperparameters} report the training D-loss and resulting MMD$^2$; MMD$^2$ $\rightarrow 0$ indicates the joint distribution of synthesized LBAs and lengths are close to the original trace---see Appendix~\ref{sec:feat_kde} Fig.~\ref{fig:llgan_feat_kde} for explicit visual comparisons.

The red dashed curves appear in Fig.~\ref{fig:real_vs_param} show LLGAN-generated HRCs. Notably, accurately reproducing LBA and length distributions does not imply HRC-fidelity. E.g., the synthetic w44 yields the closest HRC to the original despite a high MMD$^2$, whilst the synthetic w11 has the lowest MMD$^2$ but produces a clearly mismatched HRC. 

\subsection{Creating a Full Spectrum of "What-if" Traces}
\label{sec:configurability}
  \begin{table}
    \centering
    \caption{Random types supported by available tools. Only synthetic I/O generation is shown here, a tiny fraction of their full feature set.}
    \vspace{0.5em}
    \resizebox{\columnwidth}{!}{%
    \begin{tabular}{|c|*{6}{c|}*{2}{c|}c|}
      \hline
      \multirow{2}{*}{\textbf{Tool}}
        & \multicolumn{6}{c|}{\textbf{Independent}}
        & \multicolumn{2}{c|}{\textbf{Dependent}}
        & \multirow{2}{*}{\textbf{Mixed}} \\
      \cline{2-9}
        & zipf & pareto & zoned & normal & unif. & emp. & seq. & IRD & \\
      \hline
      \textbf{IOzone}~\cite{iozone}
        & \xmark & \xmark & \xmark & \xmark & \xmark & \xmark
        & \xmark & \xmark & \xmark \\ \hline
      \textbf{Bonnie++}~\cite{bonnie}
        & \xmark & \xmark & \xmark & \xmark & \xmark & \xmark
        & \xmark & \xmark & \xmark \\ \hline
      \textbf{IOmeter}~\cite{iometer}
        & \xmark & \xmark & \cmark & \xmark & \xmark & \xmark
        & \xmark & \xmark & \xmark \\ \hline
      \textbf{Sysbench}~\cite{sysbench}
        & \cmark & \cmark & \cmark & \xmark & \xmark & \xmark
        & \xmark & \xmark & \xmark \\ \hline
      \textbf{FIO}~\cite{fio}
        & \cmark & \cmark & \cmark & \cmark & \cmark & \xmark
        & \cmark & \xmark & \xmark \\ \hline
      \textbf{2DIO}
        & \cmark & \cmark & \cmark & \cmark & \cmark & \textcolor{green}{\cmark}
        & \cmark & \textcolor{green}{\cmark} & \textcolor{green}{\cmark} \\
      \hline
    \end{tabular}%
    }
    \label{tab:io_types_tools}
  \end{table}
  \begin{figure*}[h!]
    \centering
    
    \subfigure[Varying $f$ to configure HRC \emph{cliff} and \emph{plateau} positions]
    {
      \includegraphics[width=2\columnwidth, height=0.42\columnwidth]{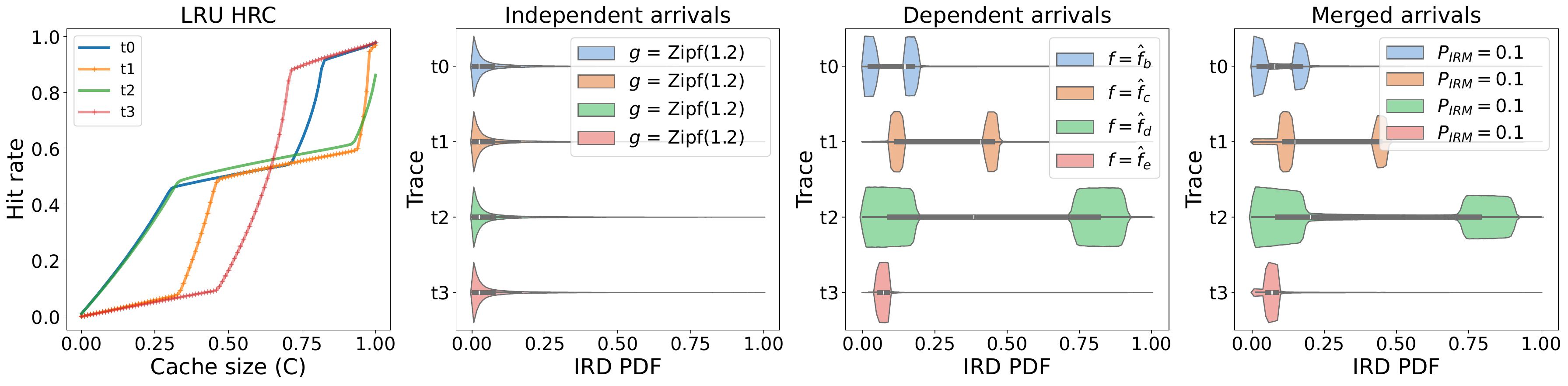}
      \label{fig:violin_f}
    }
  
    \subfigure[Varying $g$ to configure concave HRC shapes]
    {
        \includegraphics[width=2\columnwidth, height=0.42\columnwidth]{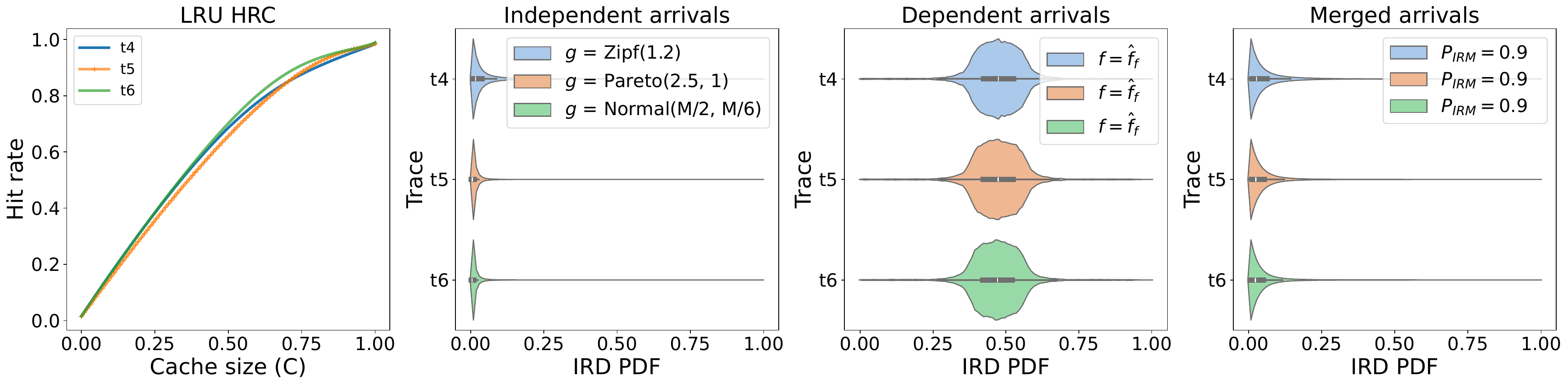}
        \label{fig:violin_g}
    }
  
    \subfigure[Varying $P_{IRM}$ to configure HRC concavity level]
    {
        \includegraphics[width=2\columnwidth, height=0.42\columnwidth]{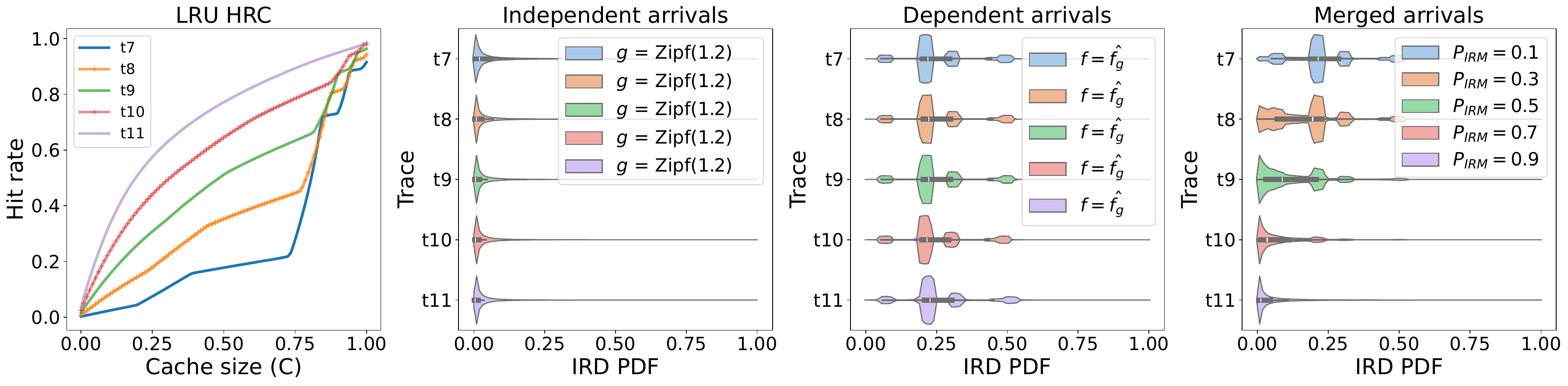}
        \label{fig:violin_pirm}
    }
    \caption{
    Effect of varying each parameter on the resulting HRCs, shown with example 2DIO-generated traces for $M=10\text{k}$ and $N=1$m. The three violin plots, from left to right, depict: (1) the IRD PDF under IRM’s independent arrivals; (2) the IRD PDF from dependent arrivals sampled via the input $f$; and (3) the combined IRD PDF obtained by merging both processes.}
    \label{fig:violin}
  \end{figure*}

  \begin{figure*}[ht!]
    \centering
    \subfigure[Scaling $M$ and $N$ with fixed $N/M$]{
        \includegraphics[width=0.5\textwidth, height=0.25\textwidth]{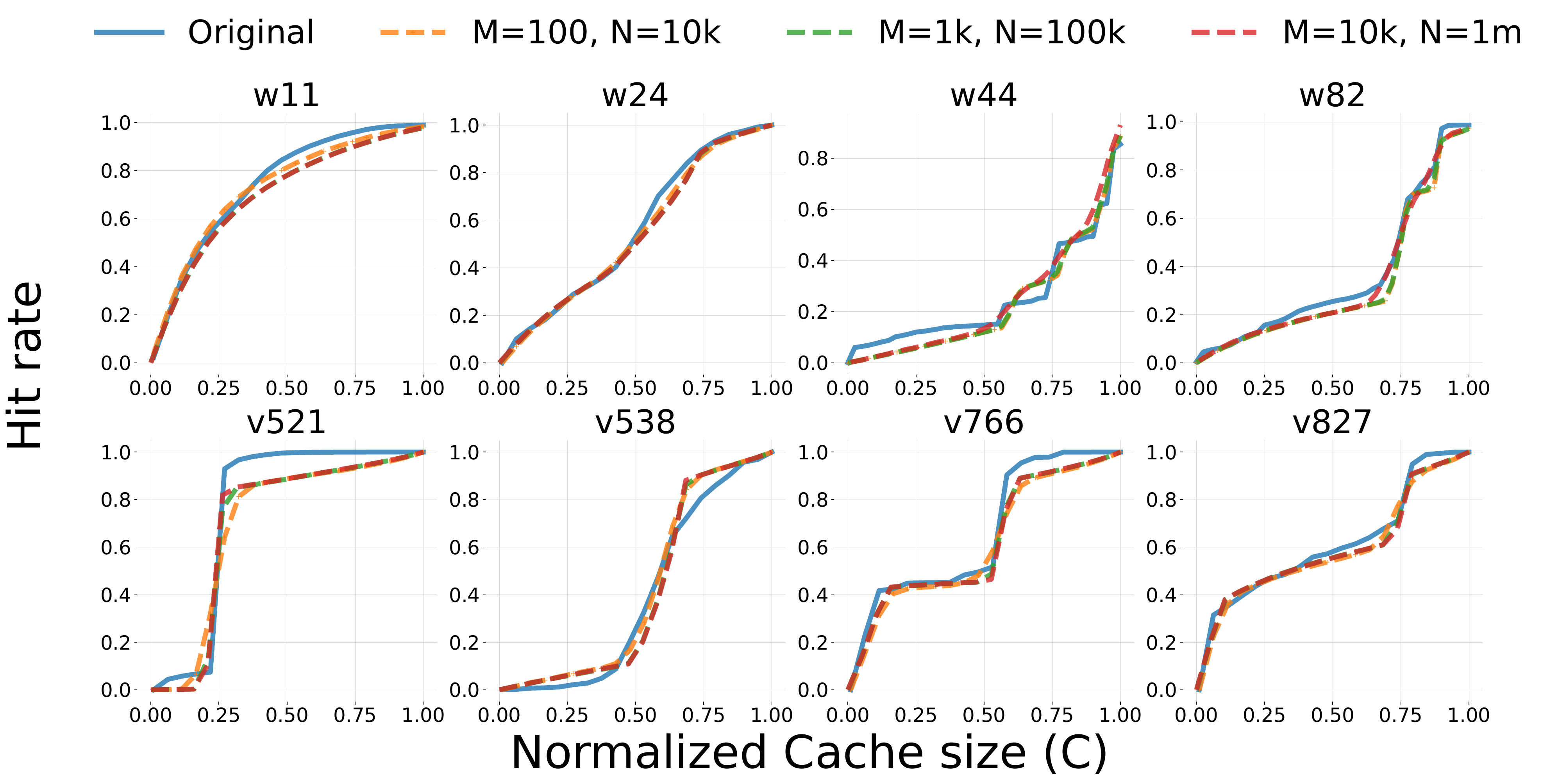}
        \includegraphics[width=0.5\textwidth, height=0.25\textwidth]{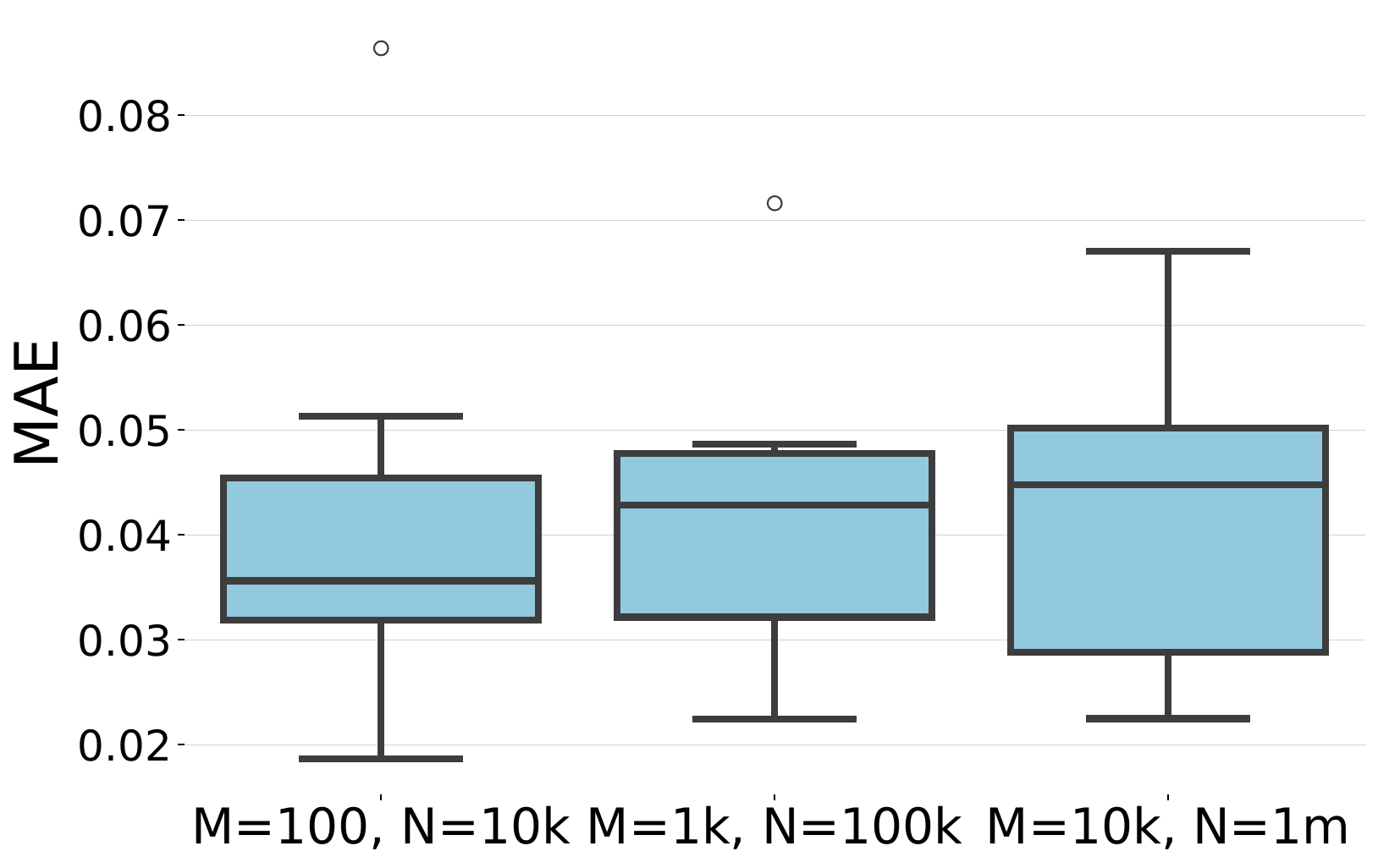}
        \label{fig:scale_m_n}
    }
    \subfigure[Scaling $M$ with fixed $N$]{
        \includegraphics[width=0.5\textwidth, height=0.25\textwidth]{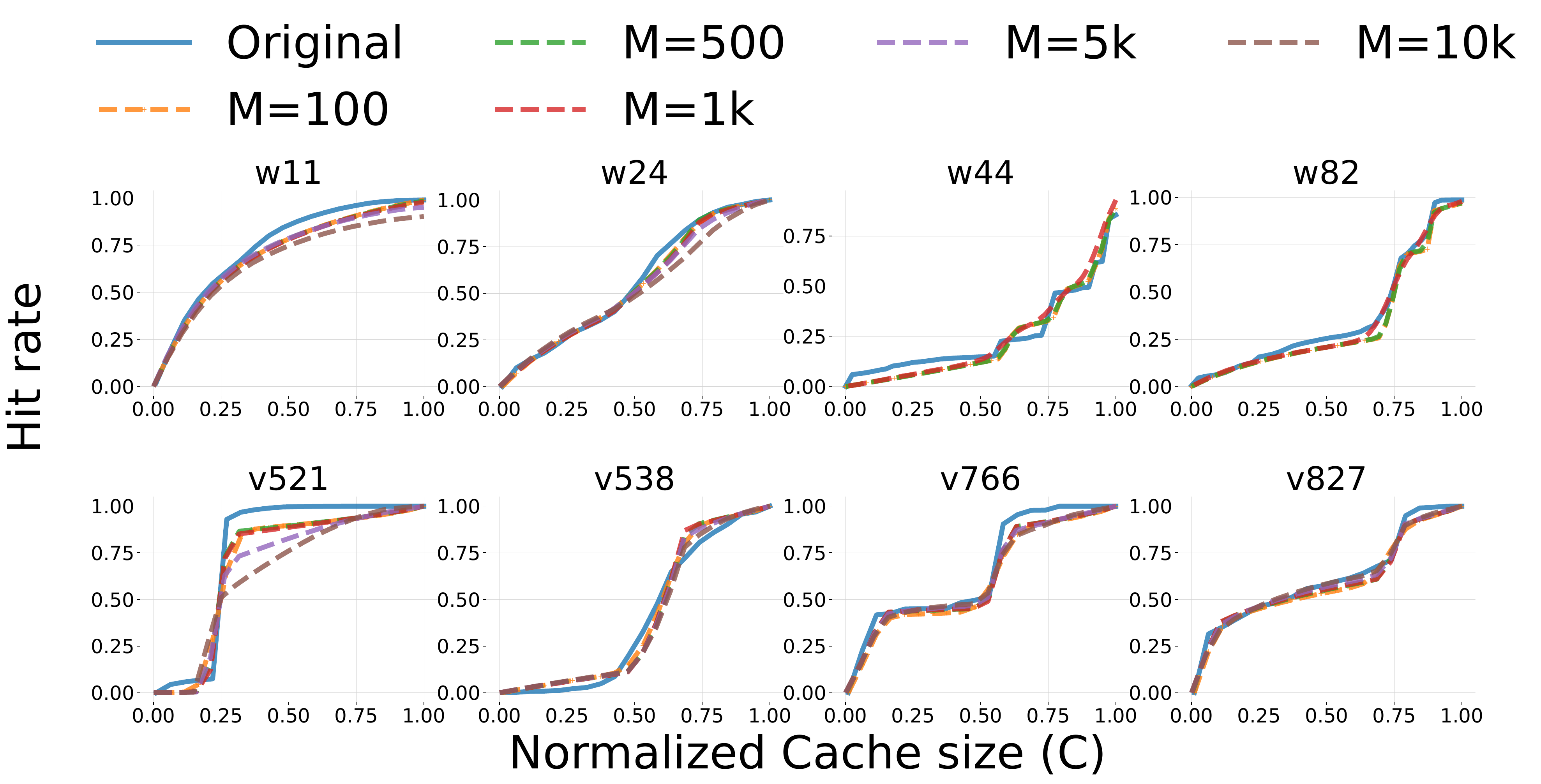}
        \includegraphics[width=0.5\textwidth, height=0.25\textwidth]{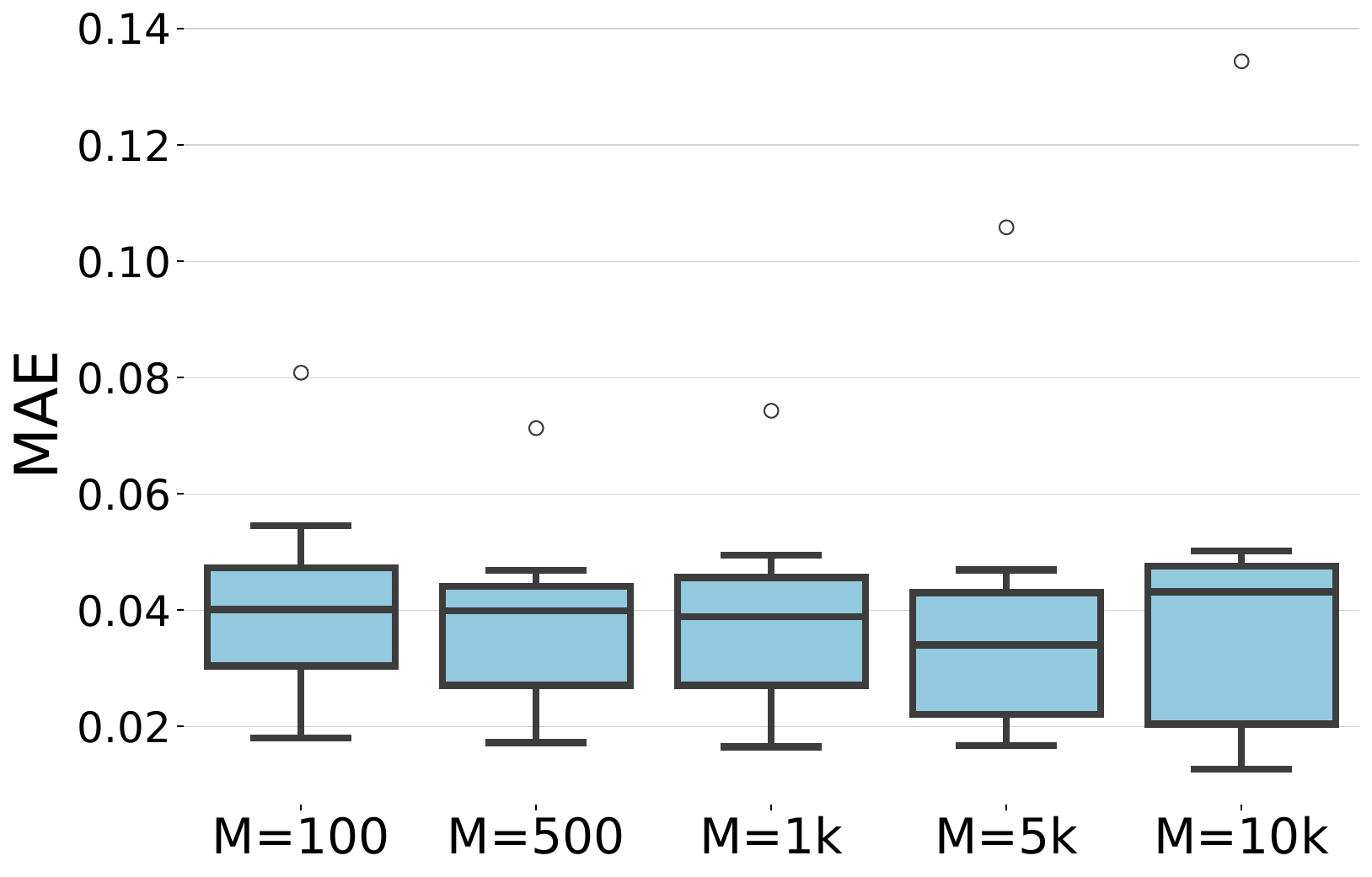}
        \label{fig:scale_m}
    }
    \subfigure[Scaling $N$ with fixed $M$]{
        \includegraphics[width=0.5\textwidth, height=0.25\textwidth]{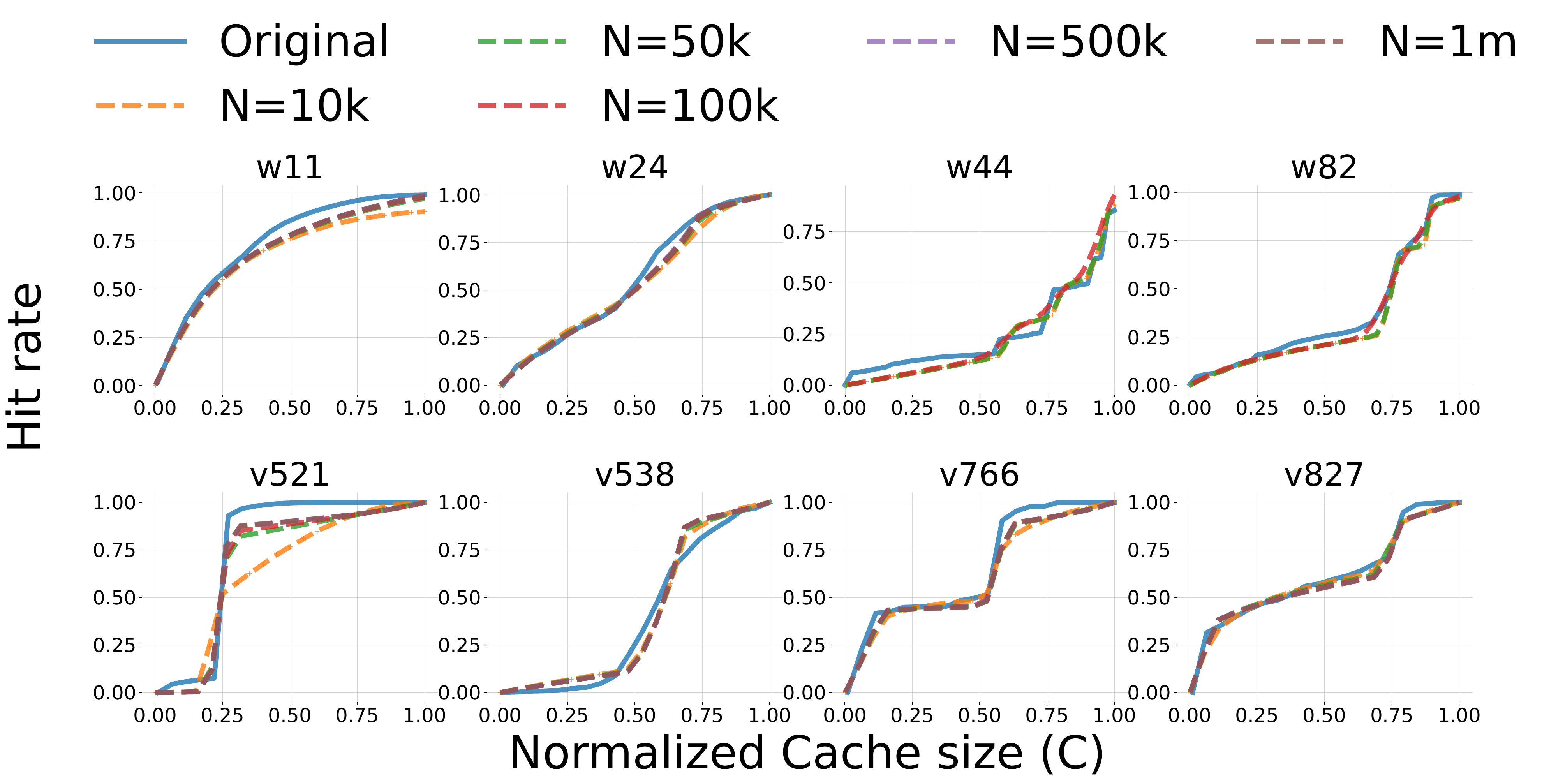}
        \includegraphics[width=0.5\textwidth, height=0.25\textwidth]{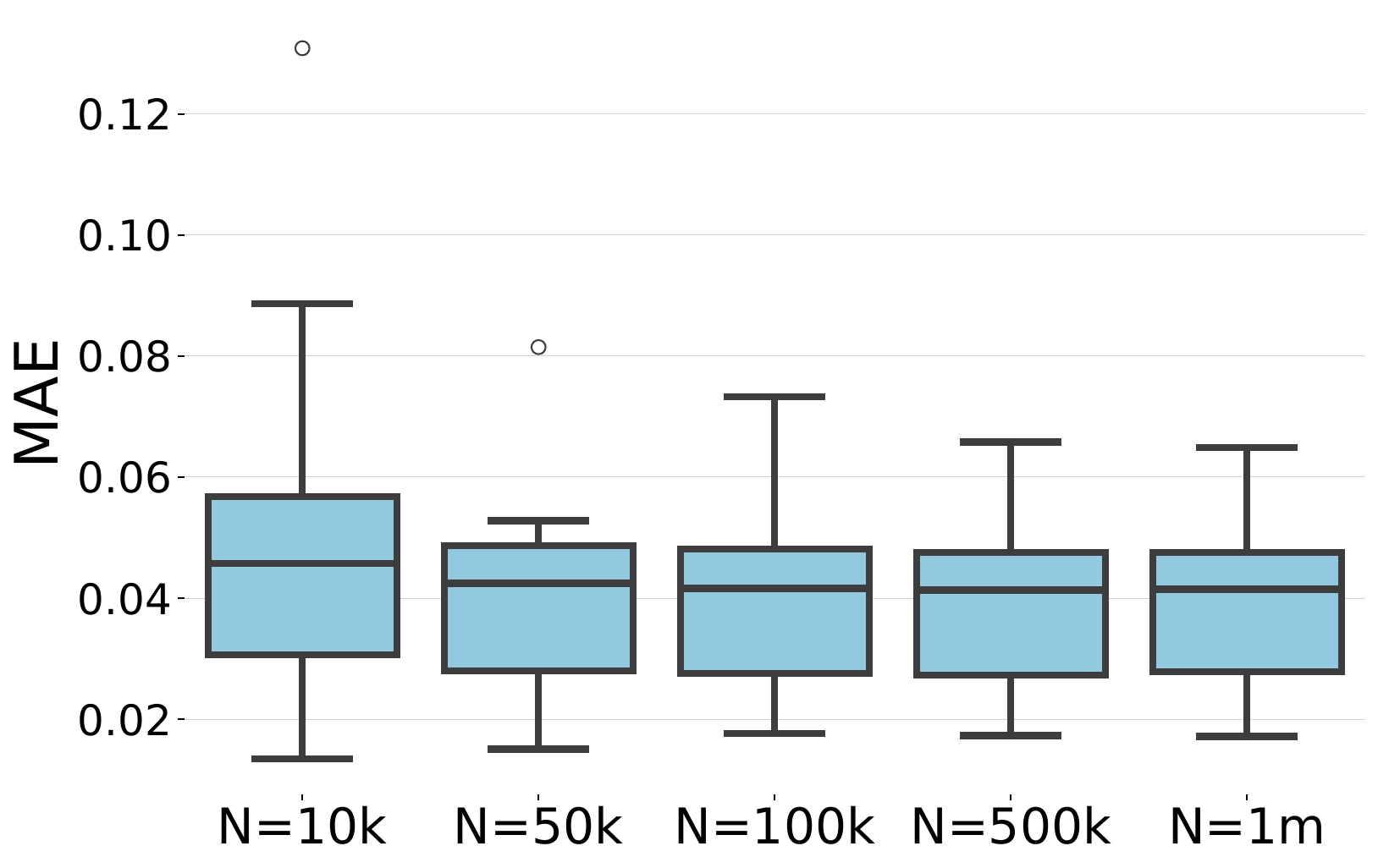}
        \label{fig:scale_n}
    }
\caption{Scalability test results. HRC accuracy under various scales of $M$ and $N$ over metric MAE.}
\end{figure*}
Typical microbenchmarks \cite{iozone, fio, bonnie, iometer, sysbench, Filebench} model workloads primarily through a few coarse knobs---read/write ratio, object size, or broader "workload personalities" such as database, file, or web servers. The cache sees whatever spatio-temporal locality naturally arises from those aggregate knobs, making the workloads effectively cache-agnostic at block level.
Few of them~\cite{fio,iometer,sysbench} allow IRM block frequency specifications, e.g., Zipf, Pareto, or customized Zoned distributions. Table~\ref{tab:io_types_tools} summarizes the random types supported by each tool, mostly limited to independent access patterns.\footnote{fio and 2DIO support heterogeneous request sizes (e.g., multi-block), inducing sequential access and triggering scan-like cache behavior, though scan locations remain intractable}

A key advantage of 2DIO is its ability to tune the full spectrum of parameters, yielding a continuum of workload behaviors. To evaluate this, we use $\hat{f}_b$ to $\hat{f}_g$\footnote{$\hat{f}g =$ \textbf{fgen}$((54, [5, 11, 12, 13, 14, 17, 30, 50], 1e-2))$; \emph{spike} values are manually re-distributed.} (see Sec.~\ref{sec:ird_sampler}) to generate 12 example traces {t0}--{t11}, each with footprint $M=10$k and length $N=1$m. 
Fig.~\ref{fig:violin} shows a separated view of the IRDs for independent, dependent, and merged arrivals, alongside the resulting HRCs. Each sub-figure illustrates how modulating $f$, $g$, and $P_{IRM}$ individually affects the simulated HRCs.
Fig.~\ref{fig:violin_f} shows how shifting \emph{spike} positions in $f$ causes corresponding changes in \emph{cliff} and \emph{plateau} positions. With fixed $P_{IRM}=0.1$, IRD \emph{spikes} primarily dictate the HRC \emph{cliff} while the IRM traffic has minor influence.
Fig.~\ref{fig:violin_g} shows how switching IRM types $g$ affects the HRC shapes. Since 90\% of arrivals are independent, $\hat{f}_f$ incurs only minor influence on the merged IRDs, yielding exponential-like merged IRDs and concave HRCs across all traces.
Fig.~\ref{fig:violin_pirm} shows the effects of adjusting $P_{IRM}$, where {t7}--{t11} are generated with fixed $g = \text{Zipf}(1.2)$ and $f = \hat{f}g$; gradually increasing $P_{IRM}$ from $0.1$ to $0.9$ results in progressively higher HRC concavity.

\subsection{Fidelity-Persistant Up- and Down-scaling}
\label{sec:scalability}
For this evaluation we regenerate each trace at various scales under fixed $\theta$ (see Sec.~\ref{sec:fidelity} Table~\ref{tab:theta_values}), assessing the HRC accuracy in terms of mean absolute error (MAE) at each scale against the original.

\noindent \textbf{Scaling $N$ and $M$ simultaneously.} We scale $M$ down from $10$k to $100$ and $N$ from $1$m to $10$k, maintaining a fixed $N/M$ ratio of $100$. Fig.~\ref{fig:scale_m_n} presents the results. The HRCs generated by 2DIO at three scales are compared with the original one. MAEs are consistently at around $0.03$ to $0.05$. While the resolution of \emph{cliffs} is "smoothed out" at the smallest scale ($M=100$, $N=10$k), the overall cache behavior persists.

\noindent \textbf{Scaling footprint $M$.}
With $N$ fixed at $100\text{k}$, varying $M$ from $10\text{k}$ to $100$ produces results in Fig.~\ref{fig:scale_m} where MAEs are found to be between $0.02$ and $0.05$.

\noindent \textbf{Scaling trace length $N$.}
Here $M$ is fixed at $1\text{k}$. Fig.~\ref{fig:scale_n} shows the HRCs and MAEs as $N$ varies from $10$k to $1$m; MAEs are around 0.04.

Observe that MAE does not necessarily improve/worsen with larger/smaller scales. Accuracy primarily depends on how well the $\theta$ parameters are tuned at the initial scales.

\subsection{Limitations}
\label{sec:limitations}
Traces generated by 2DIO follow the SPC format~\cite{spc_traces} and can be replayed on any storage systems. Users can specify a read/write ratio and a request size distribution.  
However, specifying a request size distribution that mixes single- and multi-block references effectively introduces sequential accesses at arbitrary locations, which may distort the IRD sample space and render the crafted \emph{spikes} intractable.

\section{Related Work}
\label{sec:related_works}

\noindent \textbf{Re-scaled simulations.}
A closely related field are scaling down simulations proposed by Waldspurger et al.~\cite{waldspurger_efficient_2015,waldspurger_cache_2017}, which approximates the LRU HRC by sampling requests from a given trace while preserve its underlying stack distance distribution. 

The fundamental difference between these works and 2DIO is that they translate a trace to an HRC; 2DIO does the opposite, translating an HRC to trace(s). The similarities between the two are due to both relying on prior cache modeling works going back decades (Mattson et al., 1970~\cite{mattson_evaluation_1970}; Denning, 1968~\cite{denning_working_1968}).

Scaling up synthetic traces enhances debugging and performance testing by extending trace length while preserving key distributions, as in TraceRaR~\cite{li2017tracerar}. More recently, scaling up has enabled distributed storage evaluations on single nodes or small clusters by combining down-sampled traces from multiple storage units~\cite{phothilimthana2024thesios}, creating representative or counterfactual traces based on disk capacity and placement policies.

\noindent\textbf{Workload Synthesis for CDNs.}
There is a stream of workload generation tools~\cite{sundarrajan2017footprint, sabnis2021tragen, sabnis2022jedi} used in production CDNs. The most recent JEDI~\cite{sabnis2022jedi} produces a synthetic trace that simultaneously matches the object-level properties (object size distribution, popularity distribution, request size distribtuion) and the cache-level properties (object- and byte-level HRCs) of the original trace. Their evaluations report small HRC simulation errors on workload regenerations across various policies, while all reported HRCs appear to indicate IRM-like workloads.

% %---------------------------

%-------------------------------------------------------------------------------

\section{Conclusion}
\label{sec:conclusion}
2DIO offers a practical way to generate synthetic traces that are both realistic and predictable.  
By combining short-term recency and long-term frequency characteristics in a concise parameter triplet, it produces traces with complex, \emph{non-concave} hit ratio curves, including performance \emph{cliffs} and \emph{plateaus} seen in real workloads.  
The parameterization is succinct enough to be explored exhaustively in experiments, yet highly scalable to adapt to various systems under evaluation without altering cache behavior.  
It is lightweight and easily integrates into existing benchmarking tools, enabling a standardized, reproducible, and shareable platform for cross-system benchmarking.

\begin{acks}
  We thank the anonymous reviewers and our shepherd, Bhuvan Urgaonkar, for their extensive and constructive feedback. This work was made possible thanks in part to support from Red Hat, VMware/Broadcom and NSF award CNS-1910327.
\end{acks}

\bibliographystyle{ACM-Reference-Format}
\bibliography{eurosys26.bib}

@inproceedings{sen2013reuse,
  title={Reuse-based online models for caches},
  author={Sen, Rathijit and Wood, David A},
  booktitle={Proceedings of the ACM SIGMETRICS/international conference on Measurement and modeling of computer systems},
  pages={279--292},
  year={2013}
}

@article{denning_working_1968,
	title = {The working set model for program behavior},
	volume = {11},
	issn = {0001-0782, 1557-7317},
	doi = {10.1145/363095.363141},
	language = {en},
	number = {5},
	urldate = {2022-09-09},
	journal = {Communications of the ACM},
	author = {Denning, Peter J.},
	month = may,
	year = {1968},
	pages = {323--333},
}

@article{denning1972properties,
  title={Properties of the working-set model},
  author={Denning, Peter J and Schwartz, Stuart C},
  journal={Communications of the ACM},
  volume={15},
  number={3},
  pages={191--198},
  year={1972},
  publisher={ACM New York, NY, USA}
}

@inproceedings{waldspurger_efficient_2015,
	address = {Santa Clara, CA},
	title = {Efficient {MRC} {Construction} with {SHARDS}},
	isbn = {978-1-931971-20-1},
	booktitle = {13th {USENIX} {Conference} on {File} and {Storage} {Technologies} ({FAST} 15)},
	publisher = {USENIX Association},
	author = {Waldspurger, Carl A. and Park, Nohhyun and Garthwaite, Alexander and Ahmad, Irfan},
	year = {2015},
	pages = {95--110},
}

@inproceedings{waldspurger_cache_2017,
	address = {Santa Clara, CA},
	title = {Cache {Modeling} and {Optimization} using {Miniature} {Simulations}},
	isbn = {978-1-931971-38-6},
	booktitle = {2017 {USENIX} {Annual} {Technical} {Conference} ({USENIX} {ATC} 17)},
	publisher = {USENIX Association},
	author = {Waldspurger, Carl and Saemundsson, Trausti and Ahmad, Irfan and Park, Nohhyun},
	month = jul,
	year = {2017},
	pages = {487--498},
}

@article{mattson_evaluation_1970,
	title = {Evaluation techniques for storage hierarchies},
	volume = {9},
	issn = {0018-8670},
	doi = {10.1147/sj.92.0078},
	number = {2},
	journal = {IBM Systems Journal},
	author = {Mattson, R.L. and Gecsei, J. and Slutz, D. R. and Traiger, I. L.},
	year = {1970},
	pages = {78--117},
}

@inproceedings{hu_kinetic_2016,
	address = {Denver, CO},
	title = {Kinetic {Modeling} of {Data} {Eviction} in {Cache}},
	isbn = {978-1-931971-30-0},
	booktitle = {2016 {USENIX} {Annual} {Technical} {Conference} ({USENIX} {ATC} 16)},
	publisher = {USENIX Association},
	author = {Hu, Xiameng and Wang, Xiaolin and Zhou, Lan and Luo, Yingwei and Ding, Chen and Wang, Zhenlin},
	month = jun,
	year = {2016},
	pages = {351--364},
}

@inproceedings{wires_characterizing_2014,
	address = {Broomfield, CO},
	title = {Characterizing {Storage} {Workloads} with {Counter} {Stacks}},
	isbn = {978-1-931971-16-4},
	booktitle = {11th {USENIX} {Symposium} on {Operating} {Systems} {Design} and {Implementation} ({OSDI} 14)},
	publisher = {USENIX Association},
	author = {Wires, Jake and Ingram, Stephen and Drudi, Zachary and Harvey, Nicholas J. A. and Warfield, Andrew},
	month = oct,
	year = {2014},
	pages = {335--349},
}

@article{hao_che_hierarchical_2002,
	title = {Hierarchical {Web} caching systems: modeling, design and experimental results},
	volume = {20},
	issn = {0733-8716},
	doi = {10.1109/JSAC.2002.801752},
	number = {7},
	urldate = {2022-09-09},
	journal = {IEEE Journal on Selected Areas in Communications},
	author = {{Hao Che} and {Ye Tung} and {Zhijun Wang}},
	month = sep,
	year = {2002},
	pages = {1305--1314},
}

@inproceedings{willick_disk_1993,
	address = {Pittsburgh, PA, USA},
	title = {Disk cache replacement policies for network fileservers},
	isbn = {978-0-8186-3770-4},
	doi = {10.1109/ICDCS.1993.287729},
	booktitle = {[1993] {Proceedings}. {The} 13th {International} {Conference} on {Distributed} {Computing} {Systems}},
	publisher = {IEEE Comput. Soc. Press},
	author = {Willick, D.L. and Eager, D.L. and Bunt, R.B.},
	year = {1993},
	pages = {2--11},
}

@book{coffman_operating_1973,
	address = {Englewood Cliffs, N.J},
	series = {Prentice-{Hall} series in automatic computation},
	title = {Operating systems theory},
	isbn = {978-0-13-637868-6},
	publisher = {Prentice-Hall},
	author = {Coffman, E. G. and Denning, Peter J.},
	year = {1973},
}

@inproceedings{rosenblum_design_1991,
	address = {Pacific Grove, California, United States},
	title = {The design and implementation of a log-structured file system},
	isbn = {0-89791-447-3},
	booktitle = {13th {ACM} symposium on {Operating} systems principles},
	publisher = {ACM},
	author = {Rosenblum, Mendel and Ousterhout, John K.},
	year = {1991},
	pages = {1--15},
}

@inproceedings{niu2012parda,
  title={PARDA: A fast parallel reuse distance analysis algorithm},
  author={Niu, Qingpeng and Dinan, James and Lu, Qingda and Sadayappan, Ponnuswamy},
  booktitle={2012 IEEE 26th International Parallel and Distributed Processing Symposium},
  pages={1284--1294},
  year={2012},
  organization={IEEE}
}

@inproceedings{counterstacks2014,
  title={Characterizing storage workloads with counter stacks},
  author={Wires, Jake and Ingram, Stephen and Drudi, Zachary and Harvey, Nicholas JA and Warfield, Andrew},
  booktitle={11th $\{$USENIX$\}$ Symposium on Operating Systems Design and Implementation ($\{$OSDI$\}$ 14)},
  pages={335--349},
  year={2014}
}

@inproceedings{eklov2010statstack,
  title={{StatStack}: {Efficient} modeling of {LRU} caches},
  author={Eklov, David and Hagersten, Erik},
  booktitle={2010 IEEE International Symposium on Performance Analysis of Systems \& Software (ISPASS)},
  pages={55--65},
  year={2010},
  organization={IEEE}
}

@article{aho1971principles,
  title={Principles of optimal page replacement},
  author={Aho, Alfred V and Denning, Peter J and Ullman, Jeffrey D},
  journal={Journal of the ACM (JACM)},
  volume={18},
  number={1},
  pages={80--93},
  year={1971},
  publisher={ACM New York, NY, USA}
}

@article{hasslinger_scope_2023,
	title = {Scope and {Accuracy} of {Analytic} and {Approximate} {Results} for {FIFO}, {Clock}-{Based} and {LRU} {Caching} {Performance}},
	volume = {15},
	issn = {1999-5903},
	doi = {10.3390/fi15030091},
	language = {en},
	number = {3},
	journal = {Future Internet},
	author = {Hasslinger, Gerhard and Ntougias, Konstantinos and Hasslinger, Frank and Hohlfeld, Oliver},
	month = feb,
	year = {2023},
	pages = {91},
}

@inproceedings{xiang2011linear,
  title={Linear-time modeling of program working set in shared cache},
  author={Xiang, Xiaoya and Bao, Bin and Ding, Chen and Gao, Yaoqing},
  booktitle={2011 International Conference on Parallel Architectures and Compilation Techniques},
  pages={350--360},
  year={2011},
  organization={IEEE}
}

@inproceedings{hu2015lama,
  title={$\{$LAMA$\}$: Optimized locality-aware memory allocation for key-value cache},
  author={Hu, Xiameng and Wang, Xiaolin and Li, Yechen and Zhou, Lan and Luo, Yingwei and Ding, Chen and Jiang, Song and Wang, Zhenlin},
  booktitle={2015 USENIX Annual Technical Conference (USENIX ATC 15)},
  pages={57--69},
  year={2015}
}

@inproceedings{sabnis2021tragen,
  title={TRAGEN: a synthetic trace generator for realistic cache simulations},
  author={Sabnis, Anirudh and Sitaraman, Ramesh K},
  booktitle={Proceedings of the 21st ACM Internet Measurement Conference},
  pages={366--379},
  year={2021}
}

@article{van1993properties,
  title={Properties of the miss ratio for a 2-level storage model with LRU or FIFO replacement strategy and independent references},
  author={van den Berg, Jacob and Toswley, D},
  journal={IEEE Transactions on Computers},
  volume={42},
  number={4},
  pages={508--512},
  year={1993},
  publisher={IEEE}
}

@inproceedings{wang2022separating,
  title={Separating data via block invalidation time inference for write amplification reduction in $\{$Log-Structured$\}$ storage},
  author={Wang, Qiuping and Li, Jinhong and Lee, Patrick PC and Ouyang, Tao and Shi, Chao and Huang, Lilong},
  booktitle={20th USENIX Conference on File and Storage Technologies (FAST 22)},
  pages={429--444},
  year={2022}
}

@article{pletka2018management,
  title={Management of next-generation NAND flash to achieve enterprise-level endurance and latency targets},
  author={Pletka, Roman and Koltsidas, Ioannis and Ioannou, Nikolas and Tomi{\'c}, Sa{\v{s}}a and Papandreou, Nikolaos and Parnell, Thomas and Pozidis, Haralampos and Fry, Aaron and Fisher, Tim},
  journal={ACM Transactions on Storage (TOS)},
  volume={14},
  number={4},
  pages={1--25},
  year={2018},
  publisher={ACM New York, NY, USA}
}

@inproceedings{stoica2019understanding,
  title={Understanding the design trade-offs of hybrid flash controllers},
  author={Stoica, Radu and Pletka, Roman and Ioannou, Nikolas and Papandreou, Nikolaos and Tomic, Sasa and Pozidis, Haris},
  booktitle={2019 IEEE 27th International Symposium on Modeling, Analysis, and Simulation of Computer and Telecommunication Systems (MASCOTS)},
  pages={152--164},
  year={2019},
  organization={IEEE}
}

@article{kang20202r,
  title={2r: Efficiently isolating cold pages in flash storages},
  author={Kang, Minji and Choi, Soyee and Oh, Gihwan and Lee, Sang-Won},
  journal={Proceedings of the VLDB Endowment},
  volume={13},
  number={12},
  pages={2004--2017},
  year={2020},
  publisher={VLDB Endowment}
}

@article{wang2020cache,
  title={Cache what you need to cache: Reducing write traffic in cloud cache via “one-time-access-exclusion” policy},
  author={Wang, Hua and Zhang, Jiawei and Huang, Ping and Yi, Xinbo and Cheng, Bin and Zhou, Ke},
  journal={ACM Transactions on Storage (TOS)},
  volume={16},
  number={3},
  pages={1--24},
  year={2020},
  publisher={ACM New York, NY, USA}
}

@article{tan2019ibtune,
  title={ibtune: Individualized buffer tuning for large-scale cloud databases},
  author={Tan, Jian and Zhang, Tieying and Li, Feifei and Chen, Jie and Zheng, Qixing and Zhang, Ping and Qiao, Honglin and Shi, Yue and Cao, Wei and Zhang, Rui},
  journal={Proceedings of the VLDB Endowment},
  volume={12},
  number={10},
  pages={1221--1234},
  year={2019},
  publisher={VLDB Endowment}
}

@article{yuan2024csea,
  title={CSEA: a fine-grained framework of climate-season-based energy-aware in cloud storage systems},
  author={Yuan, Zhu and Lv, Xueqiang and Xie, Ping and Ge, Haojie and You, Xindong},
  journal={The Computer Journal},
  volume={67},
  number={2},
  pages={423--436},
  year={2024},
  publisher={Oxford University Press}
}

@article{iozone,
  title={Iozone filesystem benchmark},
  author={Norcott, William D},
  journal={http://www. iozone. org/},
  year={2003}
}

@article{iometer,
  title={Iometer user’s guide},
  author={Levine, David D},
  journal={Intel Server Architecture Lab},
  volume={40},
  year={1998}
}

@misc{bonnie,
  author = {R. Coker},
  title = {{B}onnie++ now at 1.03e (last version before 2.0)! --- coker.com.au},
  howpublished = {\url{https://www.coker.com.au/bonnie++/}},
  year = {2003},
  note = {[Accessed 08-07-2024]},
}

@misc{fio,
  author = {J. Axboe},
  title = {{G}it{H}ub - axboe/fio: {F}lexible {I}/{O} {T}ester --- github.com},
  howpublished = {\url{https://github.com/axboe/fio}},
  year = {2003},
  note = {[Accessed 08-07-2024]},
}

@article{sysbench,
  title={Sysbench: a system performance benchmark},
  author={Kopytov, Alexey},
  journal={http://sysbench. sourceforge. net/},
  year={2004}
}

@inproceedings{li2017tracerar,
  title={Tracerar: An i/o performance evaluation tool for replaying, analyzing, and regenerating traces},
  author={Li, Bingzhe and Toussi, Farnaz and Anderson, Clark and Lilja, David J and Du, David HC},
  booktitle={2017 International Conference on Networking, Architecture, and Storage (NAS)},
  pages={1--10},
  year={2017},
  organization={IEEE}
}

@misc{alibaba, title={Alibaba/block-traces}, url={https://github.com/alibaba/block-traces}, journal={GitHub}, publisher={Alibaba}, author={Alibaba}}

@misc{Filebench, title={Filebench/filebench: File system and storage benchmark that uses a custom language to generate a large variety of workloads.}, url={https://github.com/filebench/filebench}, journal={GitHub}, author={Filebench}}

@inproceedings{phothilimthana2024thesios,
  title={Thesios: Synthesizing Accurate Counterfactual I/O Traces from I/O Samples},
  author={Phothilimthana, Phitchaya Mangpo and Kadekodi, Saurabh and Ghodrati, Soroush and Moon, Selene and Maas, Martin},
  booktitle={Proceedings of the 29th ACM International Conference on Architectural Support for Programming Languages and Operating Systems, Volume 3},
  pages={1016--1032},
  year={2024}
}

@inproceedings{sundarrajan2017footprint,
  title={Footprint descriptors: Theory and practice of cache provisioning in a global cdn},
  author={Sundarrajan, Aditya and Feng, Mingdong and Kasbekar, Mangesh and Sitaraman, Ramesh K},
  booktitle={Proceedings of the 13th International Conference on emerging Networking EXperiments and Technologies},
  pages={55--67},
  year={2017}
}

@inproceedings{sabnis2022jedi,
  title={JEDI: model-driven trace generation for cache simulations},
  author={Sabnis, Anirudh and Sitaraman, Ramesh K},
  booktitle={Proceedings of the 22nd ACM Internet Measurement Conference},
  pages={679--693},
  year={2022}
}

@inproceedings{zhang2024accurate,
  title={Accurate Generation of I/O Workloads Using Generative Adversarial Networks},
  author={Zhang, Heyu and Yang, Zhen and Xie, Yulai and Wu, Yafeng and Li, Jiakun and Feng, Dan and Wildani, Avani and Long, Darrell},
  booktitle={2024 International Conference on Networking, Architecture and Storage (NAS)},
  pages={1--9},
  year={2024},
  organization={IEEE}
}

@misc{spc_traces,
  author       = {University of Massachusetts, Amherst, Storage Performance Council},
  title        = {SPC Traces: Storage Performance Council Traces},
  year         = {n.d.},
  howpublished = {\url{https://skulddata.cs.umass.edu/traces/storage/SPC-Traces.pdf}},
  note         = {Accessed: 2025-01-02}
}

@inproceedings{akiba2019optuna,
  title={Optuna: A next-generation hyperparameter optimization framework},
  author={Akiba, Takuya and Sano, Shotaro and Yanase, Toshihiko and Ohta, Takeru and Koyama, Masanori},
  booktitle={Proceedings of the 25th ACM SIGKDD international conference on knowledge discovery \& data mining},
  pages={2623--2631},
  year={2019}
}

@misc{ircache_2000,
  title   = {The First Semi-Annual Web Caching Cache-off},
  author  = {IRCache},
  year    = {2000},
  note    = {\url{http://www.ircache.net/n01} (archived at \url{https://web.archive.org/web/*/http://www.ircache.net/n01})},
}

\clearpage
% \newpage
\appendix
\section{2DIO Built‐in Trace Profiles}
\label{sec:default_vis}
\begin{figure}[b!]
  \centering
      \subfigure[$\theta_a=\langle P_{IRM}=1.0, g = \text{Zipf}(3.0), f= \text{None}\rangle$]{
          \includegraphics[width=0.95\columnwidth, height=0.32\columnwidth]{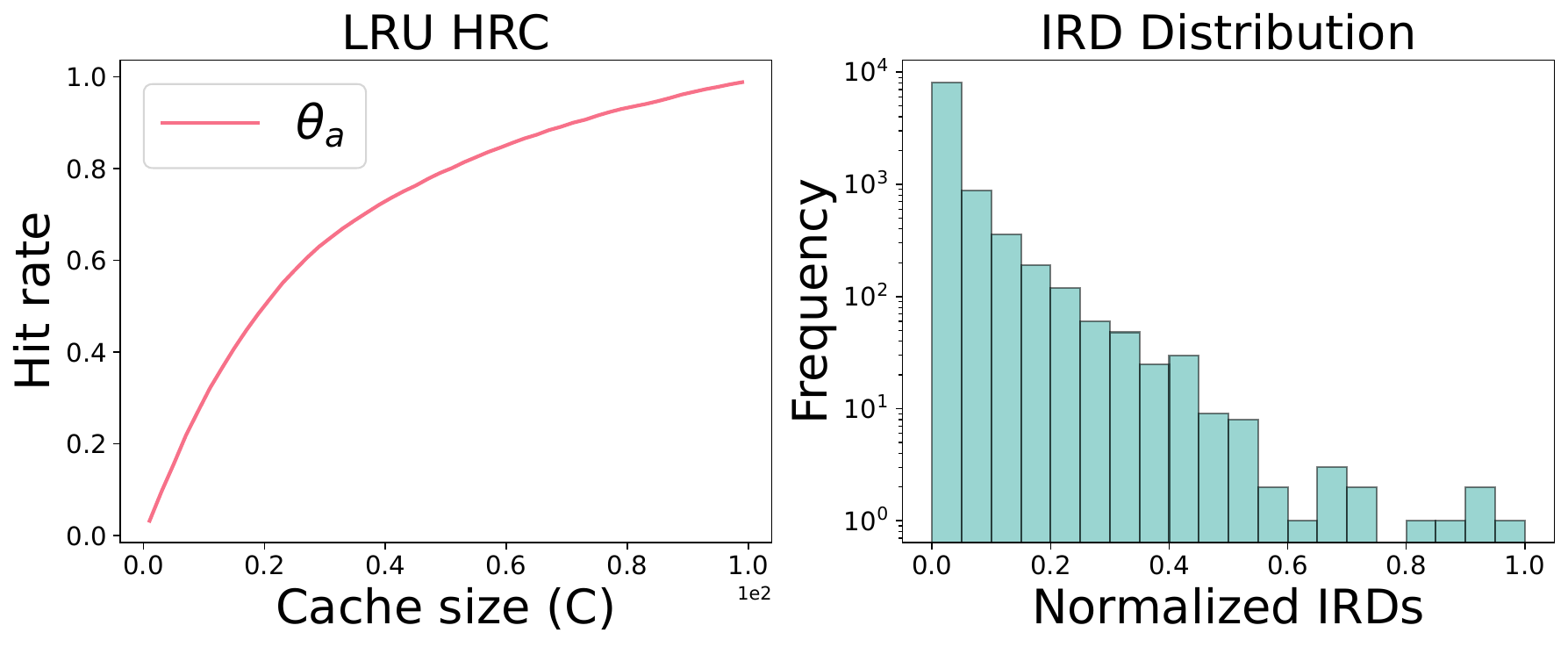 }
          \label{fig:pia_mrc_ird}
      }
      \subfigure[$\theta_b = \langle P_{IRM} = 0.0, g = \text{None}, f = \hat{f}_b \rangle$]{ 
          \includegraphics[width=0.95\columnwidth, height=0.32\columnwidth]{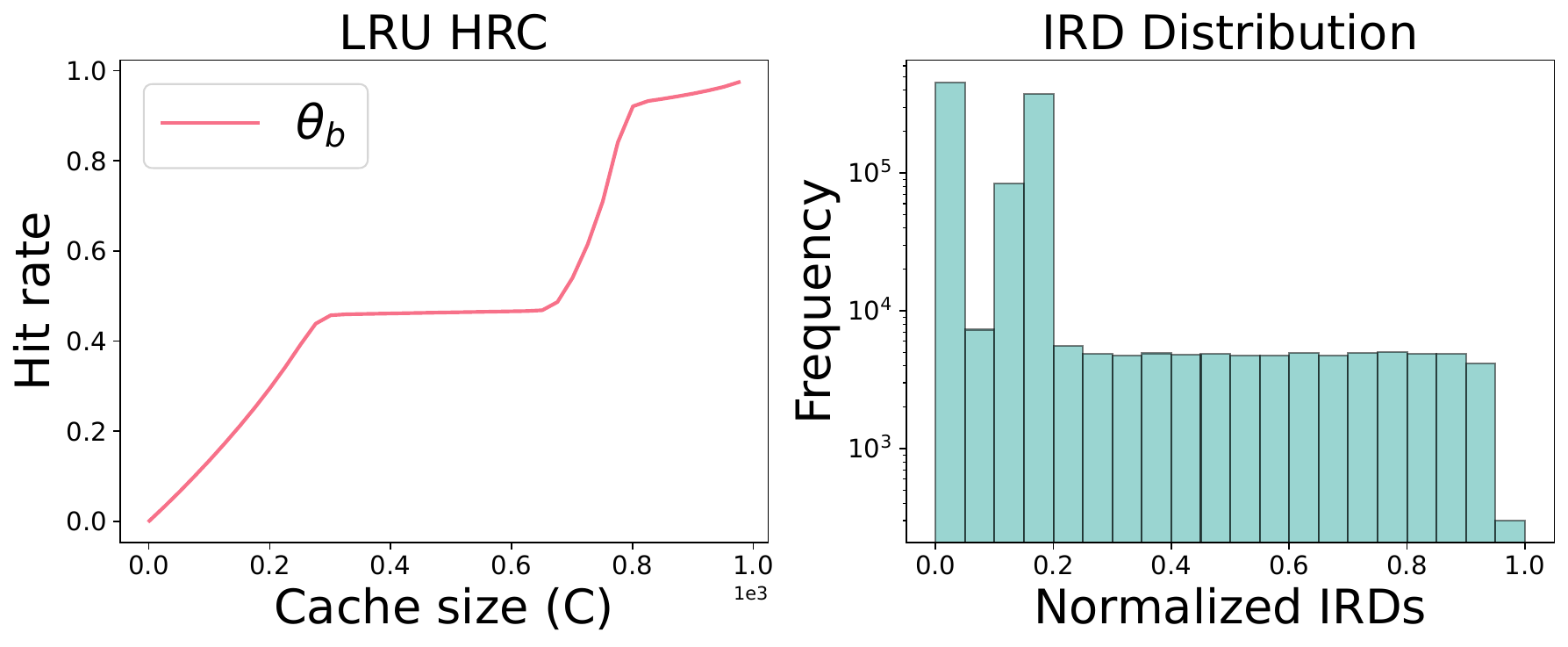 }
          \label{fig:pib_mrc_ird}
      }
      \subfigure[$\theta_c = \langle P_{IRM} = 0.0, g = \text{None}, f = \hat{f}_c \rangle$]{
          \includegraphics[width=0.95\columnwidth, height=0.32\columnwidth]{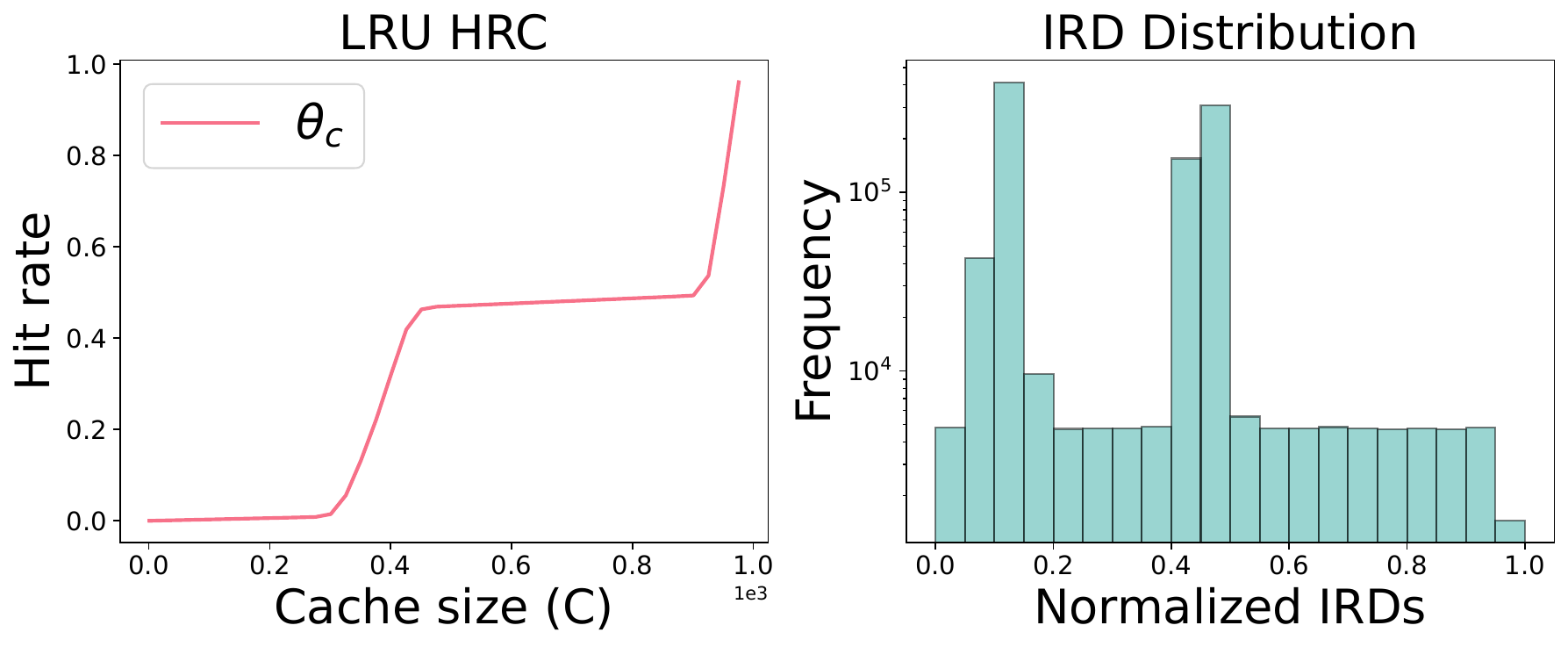 }
          \label{fig:pic_mrc_ird}
      }
      \subfigure[$\theta_d = \langle P_{IRM} = 0.0, g = \text{None}, f = \hat{f}_d \rangle$]{
          \includegraphics[width=0.95\columnwidth, height=0.32\columnwidth]{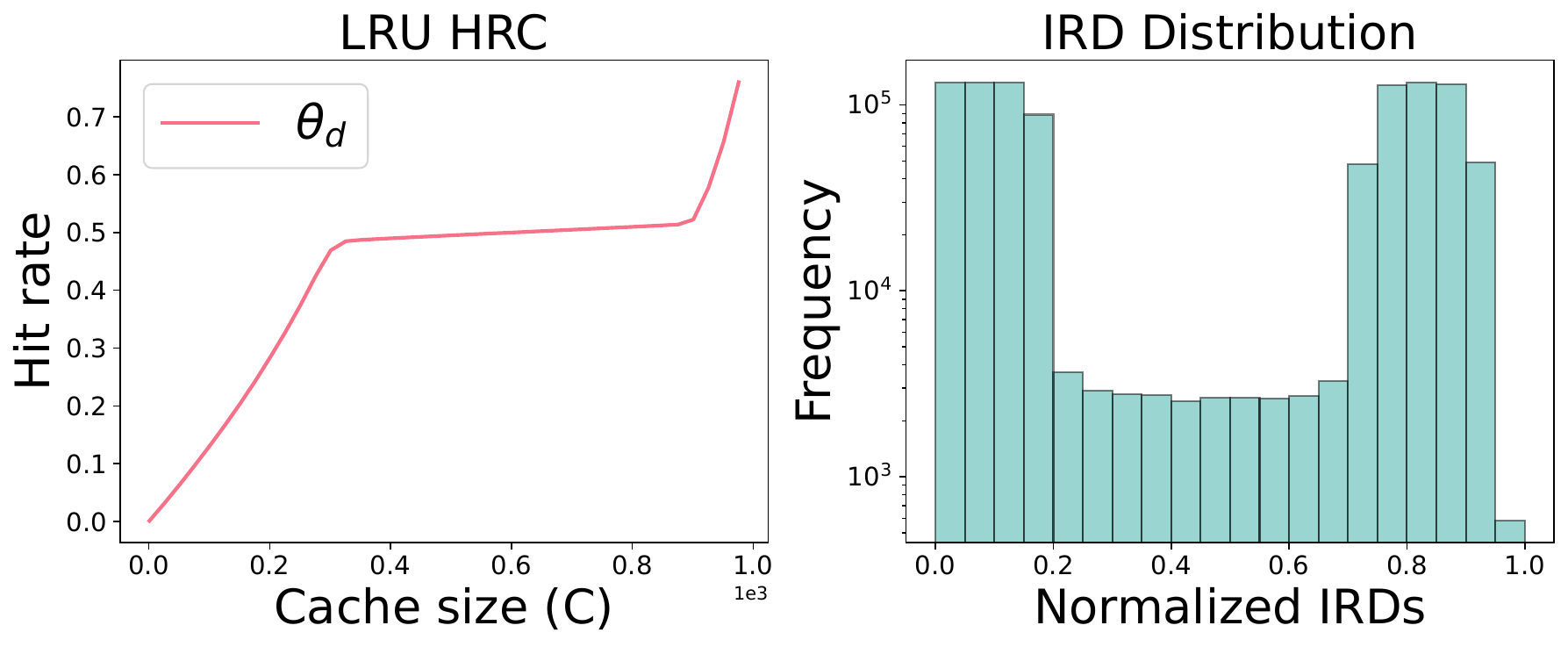 }
          \label{fig:pid_mrc_ird}
      }
      \subfigure[$\theta_e = \langle P_{IRM} = 0.0, g = \text{None}, f = \hat{f}_e \rangle$]{
          \includegraphics[width=0.95\columnwidth, height=0.32\columnwidth]{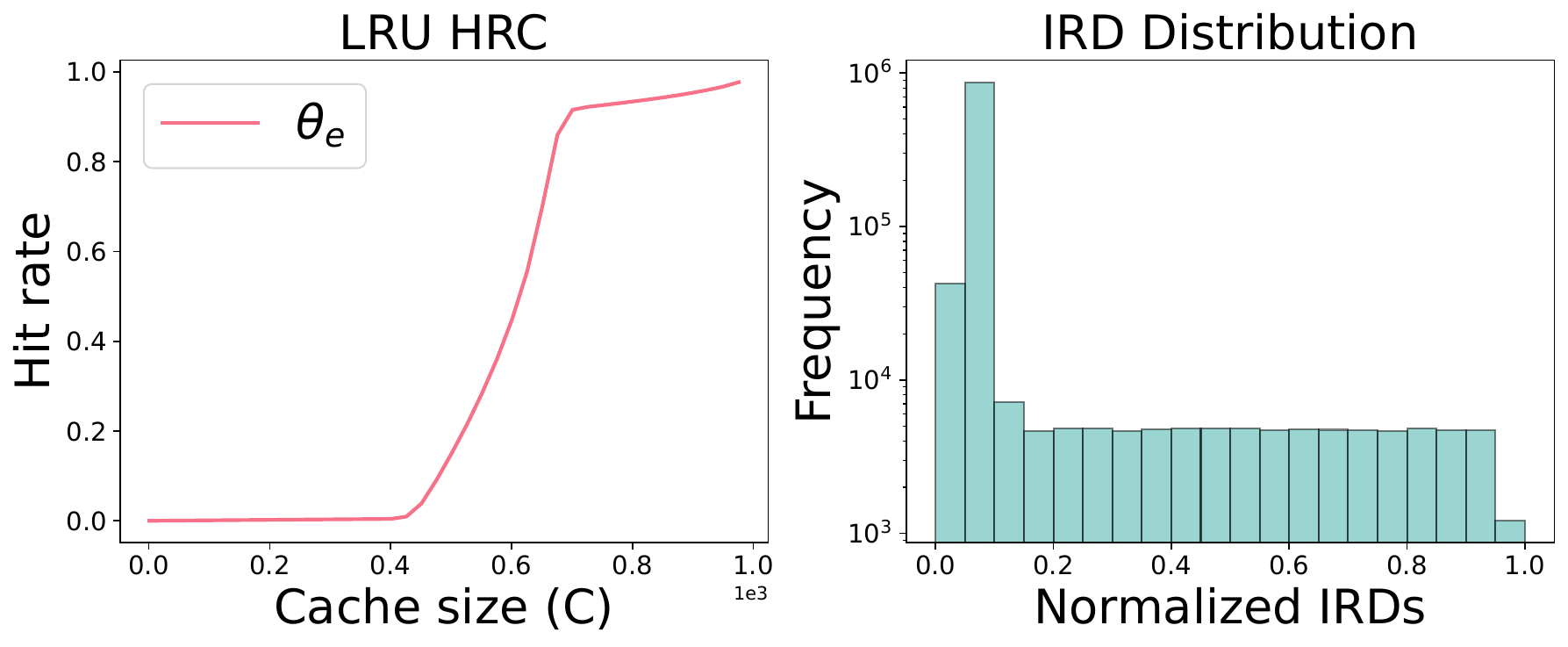 }
          \label{fig:pie_mrc_ird}
      }
      \subfigure[$\theta_f = \langle P_{IRM} = 0.0, g = \text{None}, f = \hat{f}_f \rangle$]{
          \includegraphics[width=0.95\columnwidth, height=0.32\columnwidth]{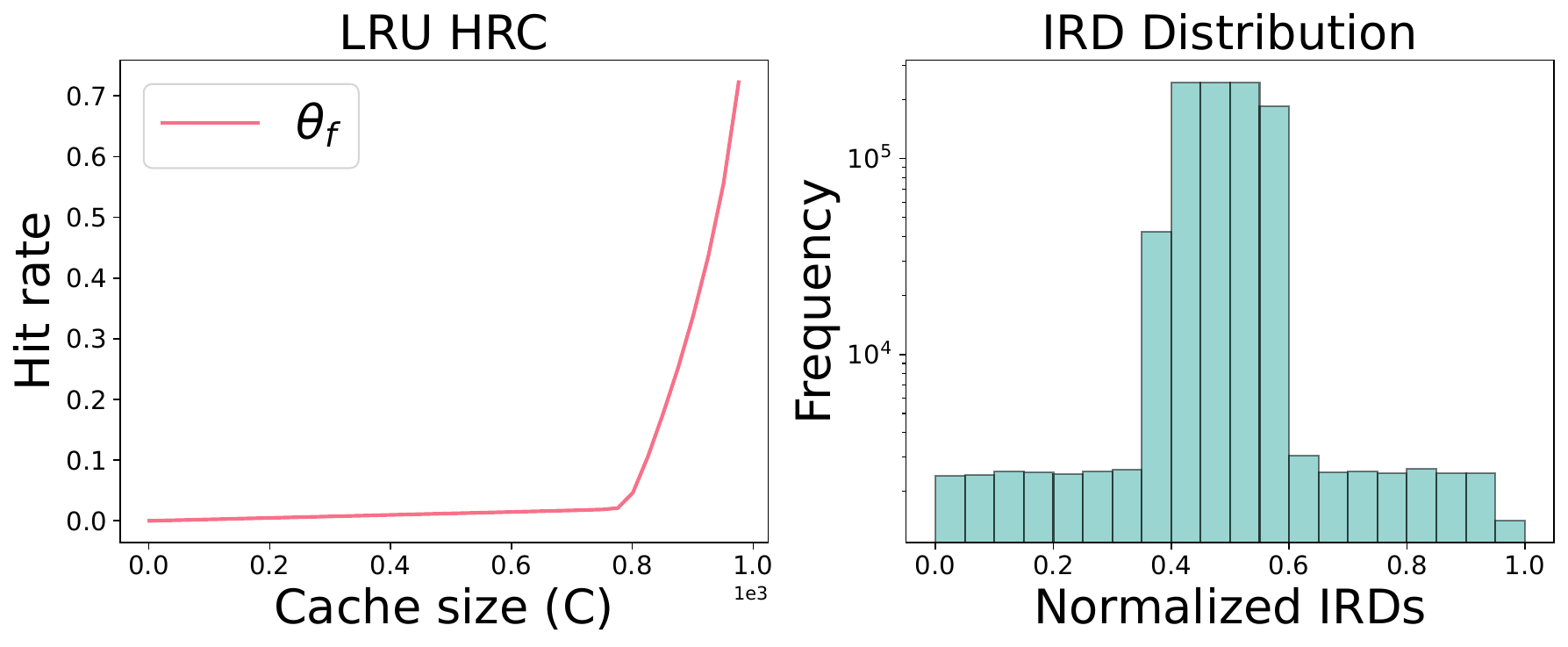 }
          \label{fig:pif_mrc_ird}
      }
    \caption{HRCs and IRD distributions for the 2DIO default trace profiles $\theta_a$–$\theta_f$.}
  \label{fig:canonical_mrc_ird}
\end{figure}

Table \ref{tab:parameters} lists the default trace profiles introduced in Sec.~\ref{sec:ird_sampler}, with \textbf{fgen} defined by Eq.~\eqref{eq:fgen}. Fig.~\ref{fig:canonical_mrc_ird} visualizes the HRC and IRD distributions for each profile, presenting a set of canonical workload behaviors that reflect commonly observed real-world patterns.
\begin{table}[H]
  \centering
  \caption{2DIO default trace profiles.}
  \vspace{0.5em}
  \begin{tabular}{l|l|l|l}
  \hline
  \textbf{} & $P_{IRM}$ & $g$ & $f$ \\
  \hline
  $\theta_a$ & 1.0 & $\text{Zipf}(3.0)$ & $\hat{f}_a = \text{None}$ \\
  $\theta_b$ & 0.0 & $\text{None}$ & $\hat{f}_b = \textbf{fgen}(20, [0, 3], 5e-3)$ \\
  $\theta_c$ & 0.0 & $\text{None}$ & $\hat{f}_c = \textbf{fgen}(20, [2, 9], 5e-3)$ \\
  $\theta_d$ & 0.0 & $\text{None}$ & $\hat{f}_d = \textbf{fgen}(5, [0, 4], 1e-2)$ \\
  $\theta_e$ & 0.0 & $\text{None}$ & $\hat{f}_e = \textbf{fgen}(20, [1], 5e-3)$ \\
  $\theta_f$ & 0.0 & $\text{None}$ & $\hat{f}_f = \textbf{fgen}(5, [2], 5e-3)$ \\
  \hline
  \end{tabular}%
  \label{tab:parameters}
\end{table}
\section{LLGAN Synthetic Traces Validation}
\label{sec:feat_kde}
\begin{figure*}[h]
  \centering
  \includegraphics[width=0.95\columnwidth, height=0.65\columnwidth]{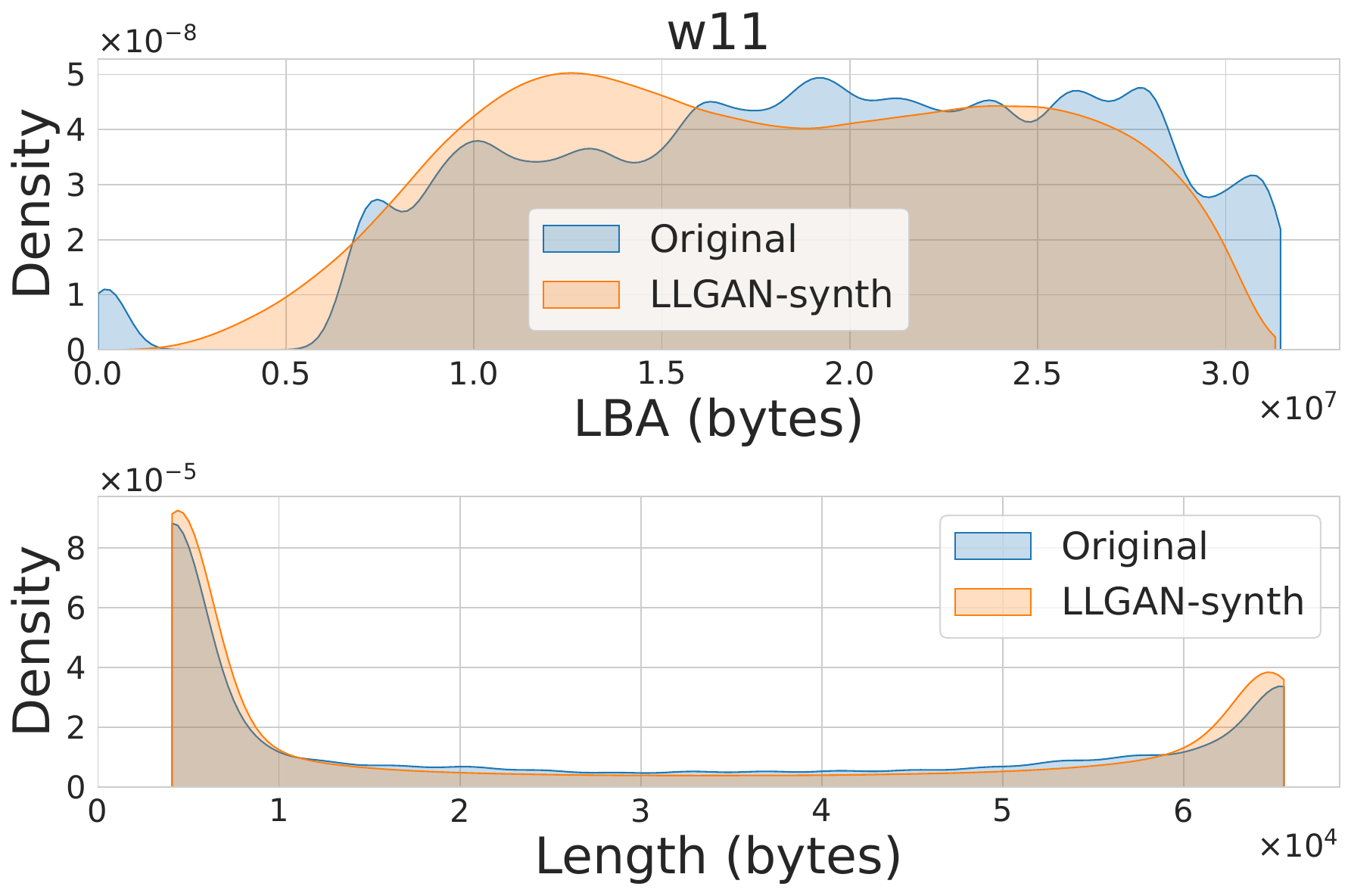}
  \vspace{0.4em}
  \includegraphics[width=0.95\columnwidth, height=0.65\columnwidth]{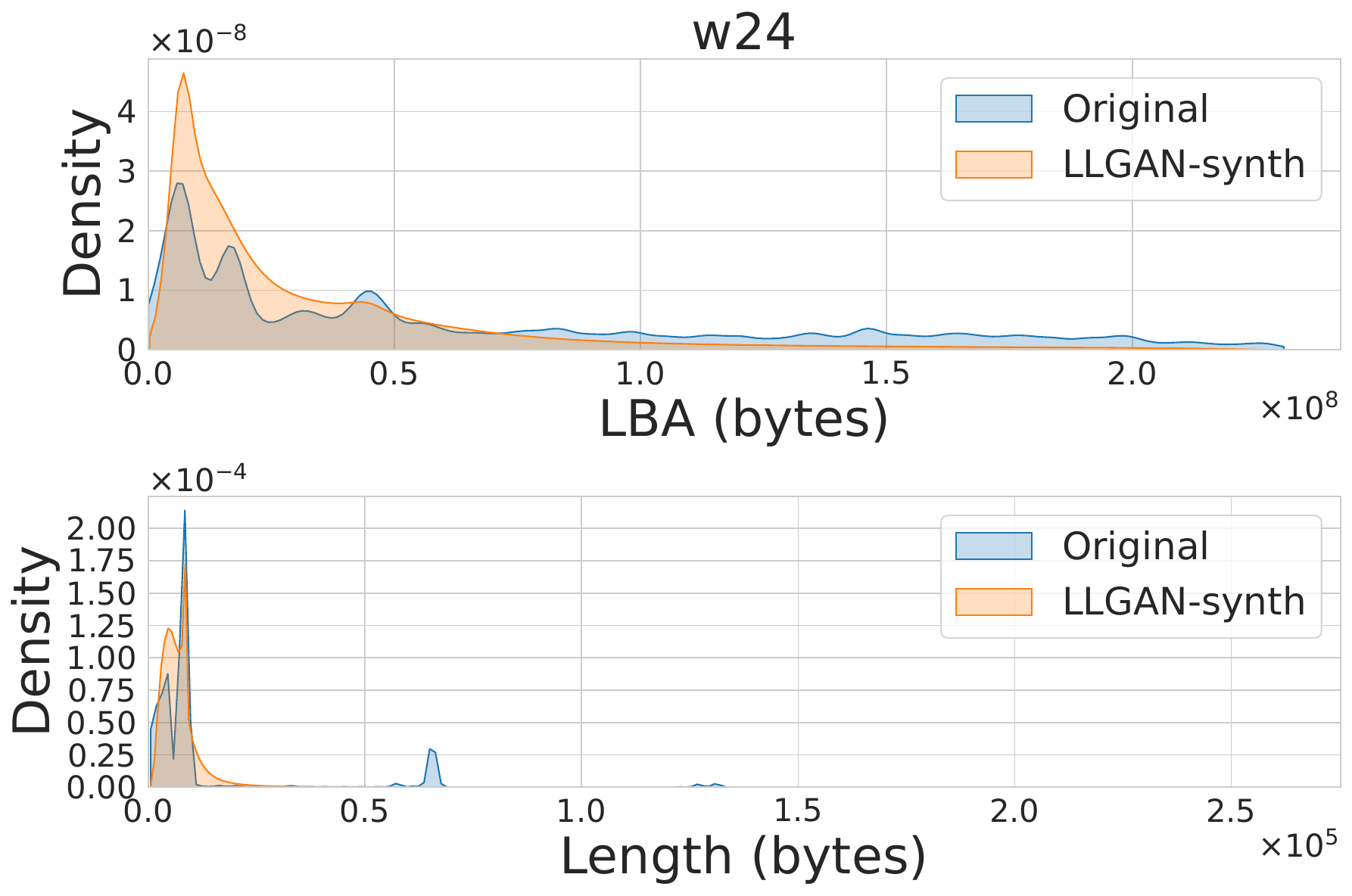}
  \vspace{0.4em}
  \includegraphics[width=0.95\columnwidth, height=0.65\columnwidth]{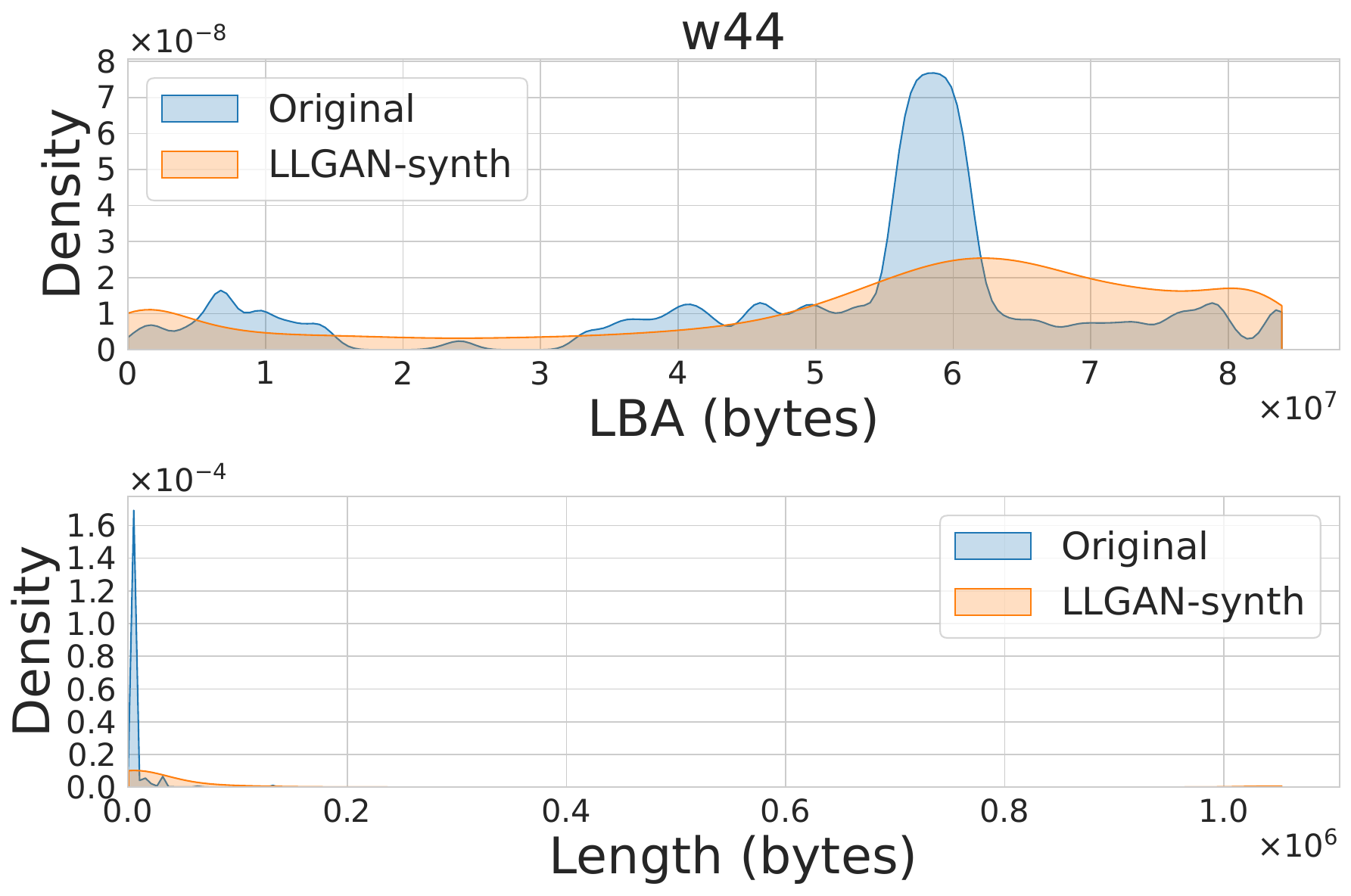}
  \vspace{0.4em}
  \includegraphics[width=0.95\columnwidth, height=0.65\columnwidth]{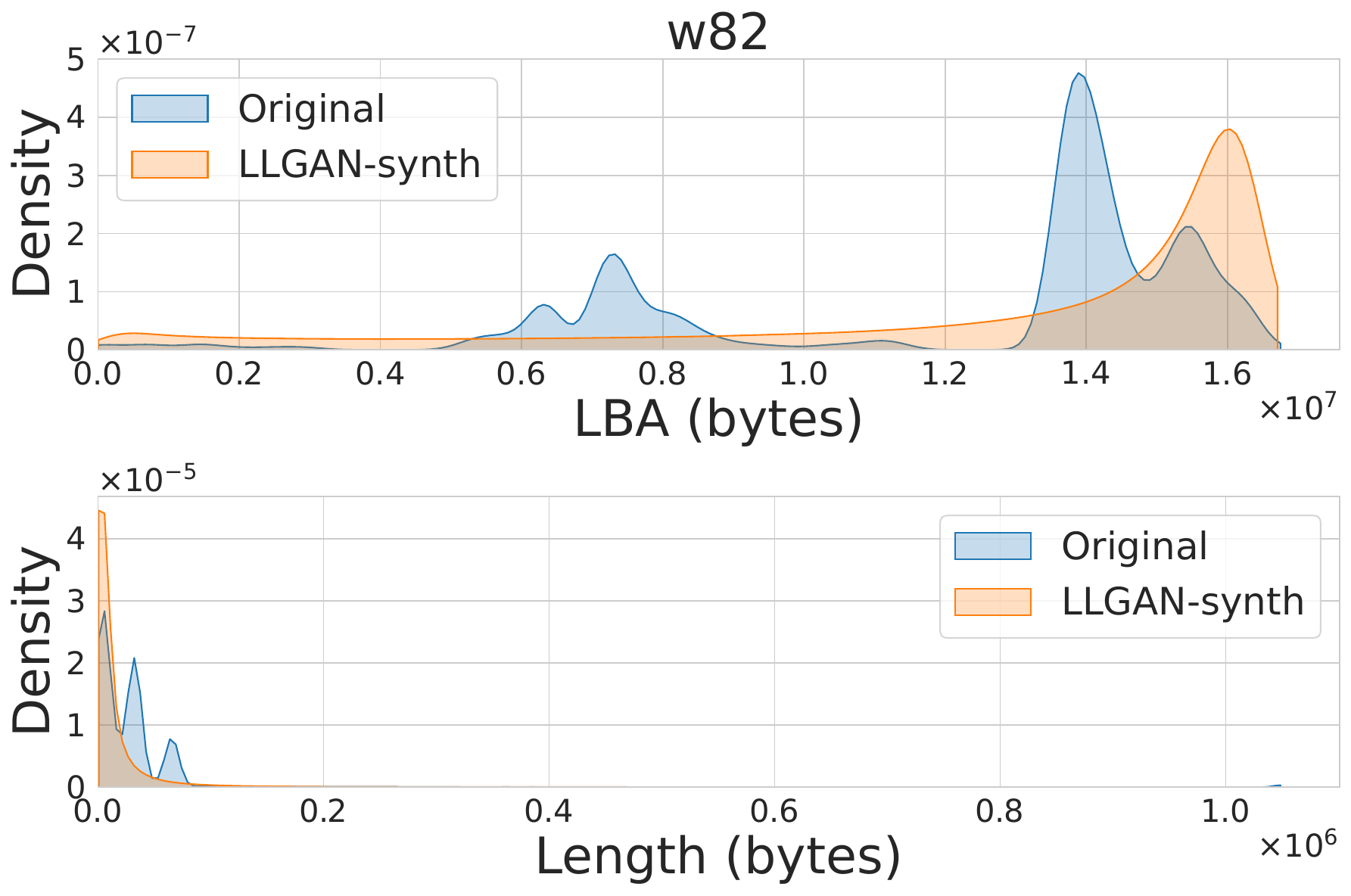}
  \vspace{0.4em}
  \includegraphics[width=0.95\columnwidth, height=0.65\columnwidth]{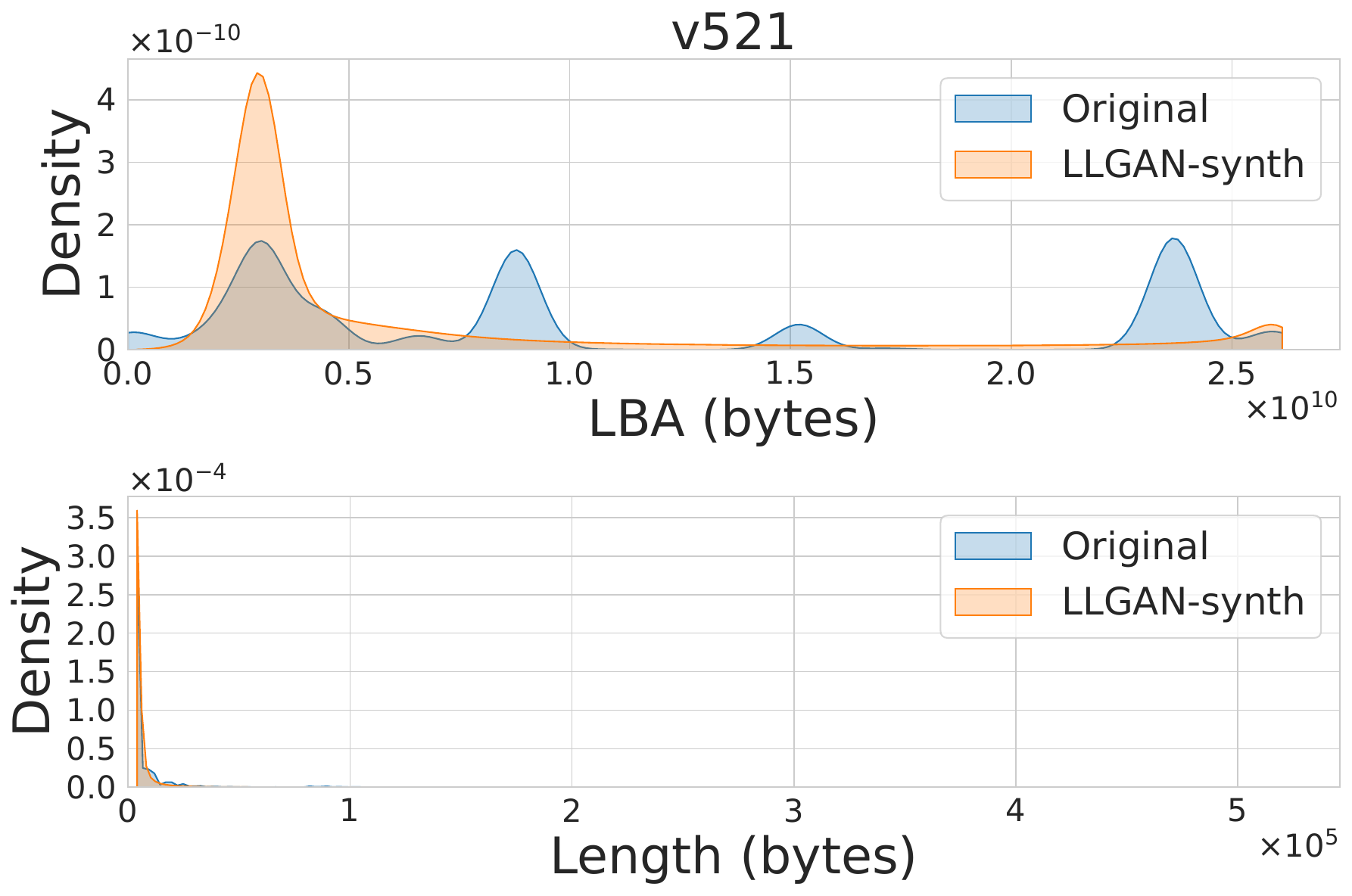}
  \vspace{0.4em}
  \includegraphics[width=0.95\columnwidth, height=0.65\columnwidth]{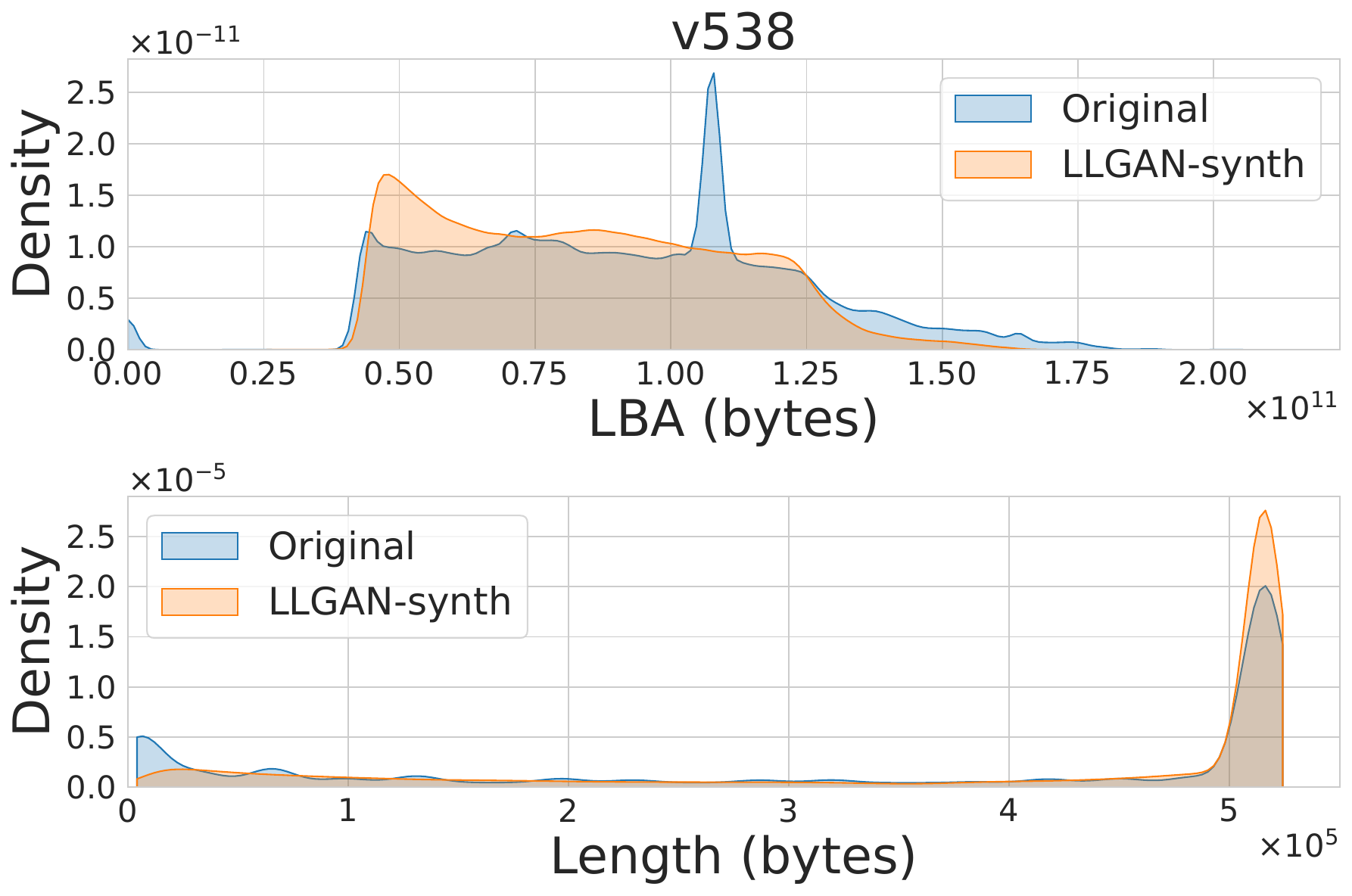}
  \vspace{0.4em}
  \includegraphics[width=0.95\columnwidth, height=0.65\columnwidth]{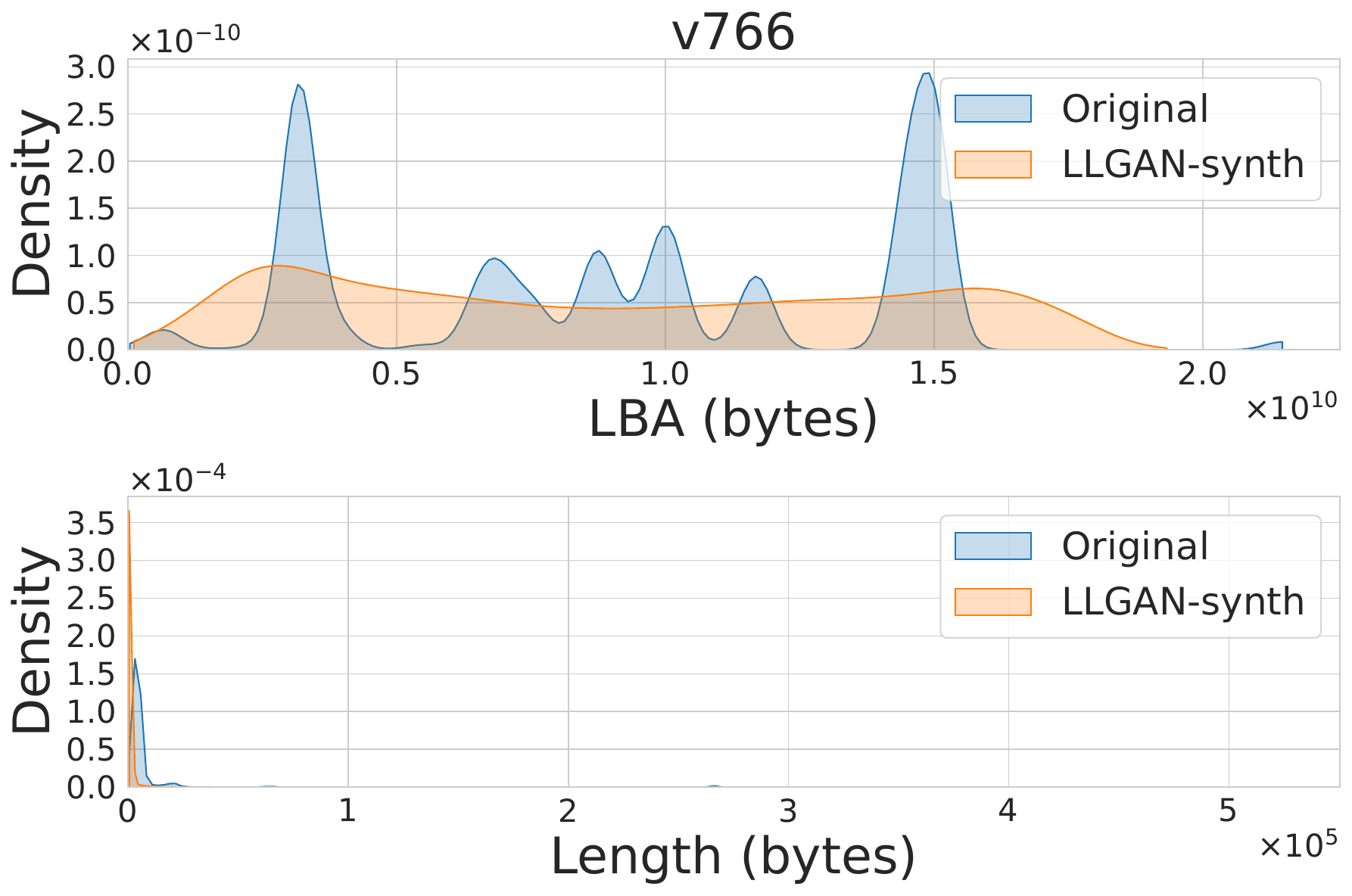}
  \vspace{0.4em}
  \includegraphics[width=0.95\columnwidth, height=0.65\columnwidth]{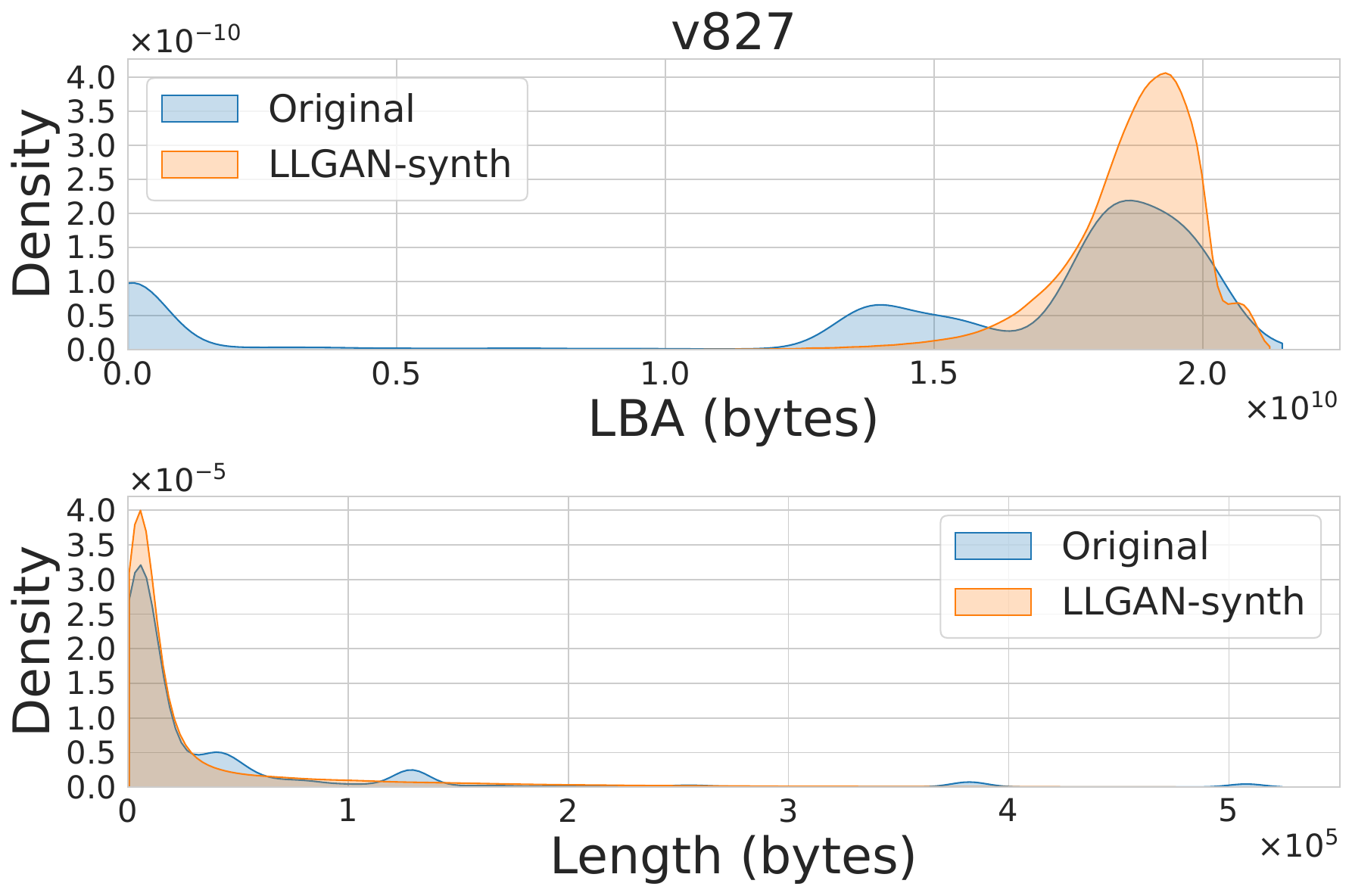}
  \caption{KDEs of LBAs and lengths, showing distributional similarity between LLGAN-synthesized traces vs. originals.}
  \label{fig:llgan_feat_kde}
\end{figure*}

Fig.~\ref{fig:llgan_feat_kde} plots the kernel density estimation (KDE) of LBA and length fidelity for LLGAN-synthesized traces against the originals as a sanity check for Sec.~\ref{sec:fidelity}. Synthetic w11 matches the originals most closely, in line with its lowest MMD$^2$ seen in Table~\ref{tab:llgan_hyperparameters}, whereas synthetic w44 shows a clear deviation (median MMD$^2 \approx 0.06$), yet produces the highest HRC fit.

\end{document}